\documentclass[gmd, manuscript]{copernicus}

\usepackage{multirow}
\usepackage{adjustbox}

\nolinenumbers
\begin{document}

\title{Quantifying and attributing time step sensitivities in 
present-day climate simulations conducted with EAMv1}

% \Author[affil]{given_name}{surname}

\Author[1]{Hui}{Wan}
\Author[1]{Shixuan}{Zhang}
\Author[1]{Philip J.}{Rasch}
\Author[2,1]{Vincent E.}{Larson}
\Author[3]{Xubin}{Zeng}
\Author[4,1]{Huiping}{Yan}

\affil[1]{Atmospheric Sciences and Global Change Division, Pacific Northwest National Laboratory}
\affil[2]{Department of Mathematical Sciences, University of Wisconsin -- Milwaukee}
\affil[3]{Department of Hydrology and Atmospheric Sciences, University of Arizona}
\affil[4]{School of Atmospheric Science, Nanjing University of Information Science and Technology}

%% The [] brackets identify the author with the corresponding affiliation. 1, 2, 3, etc. should be inserted.

%% If an author is deceased, please mark the respective author name(s) with a dagger, e.g. "\Author[2,$\dag$]{Anton}{Aman}", and add a further "\affil[$\dag$]{deceased, 1 July 2019}".

%% If authors contributed equally, please mark the respective author names with an asterisk, e.g. "\Author[2,*]{Anton}{Aman}" and "\Author[3,*]{Bradley}{Bman}" and add a further affiliation: "\affil[*]{These authors contributed equally to this work.}".

\correspondence{Hui Wan (Hui.Wan@pnnl.gov)}

\runningtitle{Time step sensitivities in EAMv1}

\runningauthor{Wan et al.}

\received{}
\pubdiscuss{} %% only important for two-stage journals
\revised{}
\accepted{}
\published{}

%% These dates will be inserted by Copernicus Publications during the typesetting process.

\firstpage{1}

\maketitle

%\noindent{\large {\it Manuscript submitted to Geoscientific Model Development (GMD)}}
%\bigskip

\begin{abstract}
This study assesses the relative importance of time integration error in present-day 
climate simulations conducted with the atmosphere component of the 
Energy Exascale Earth System Model version 1 (EAMv1) at 1-degree horizontal resolution.
We show that a factor-of-6 reduction of time step size in 
all major parts of the model leads to significant changes in the long-term mean climate. 
Examples of changes in 10-year mean zonal averages include:
(1) up to 0.5~K of warming in the lower troposphere and cooling 
in the tropical and subtropical upper troposphere, 
(2) 1\%--10\% decreases in relative humidity throughout the troposphere, and 
(3) 10\%--20\% decreases in cloud fraction in the upper troposphere and 
 decreases exceeding 20\% in the subtropical lower troposphere. 
In terms of the 10-year mean geographical distribution, systematic decreases 
of 20\%--50\% are seen in total cloud cover and cloud radiative effects
in the subtropics.
These changes imply that the reduction of temporal truncation errors 
leads to a notable although unsurprising degradation of agreement 
between the simulated and observed present-day climate;
to regain optimal climate fidelity 
in the absence of those truncation errors,
the model would require retuning.

A coarse-grained attribution of the time step sensitivities is carried out 
by shortening time steps used in various components of EAM or 
by revising the numerical coupling between some processes.
Our analysis leads to the finding that the 
marked decreases in subtropical low-cloud fraction  
and total cloud radiative effect are caused not by the step size used for the collectively subcycled
turbulence, shallow convection and stratiform cloud macro- and microphysics parameterizations
but rather by the step sizes used outside those subcycles.
Further analysis suggests that the coupling frequency between the subcycles 
and the rest of EAM significantly affects the subtropical marine stratocumulus decks while
deep convection has significant impacts on trade cumulus.
The step size of the cloud macro- and microphysics subcycle
itself appears to have a primary impact on cloud fraction
in the upper troposphere 
and also in the mid-latitude near-surface layers.
Impacts of step sizes used by the dynamical core and the radiation parameterization 
appear to be relatively small.
These results provide useful clues for future studies aiming at
understanding and addressing the root causes of sensitivities 
to time step sizes and process coupling frequencies in EAM.

While this study focuses on EAMv1, and 
the conclusions are likely model-specific,
the presented experimentation strategy 
has general value for weather and climate model development,
as the methodology can help identify and understand 
sources of time integration error in sophisticated multi-component models.
\end{abstract}

%\copyrightstatement{TEXT}

\introduction  %% \introduction[modified heading if necessary]

Atmospheric General Circulation Models (AGCMs) simulate physical and chemical processes in the Earth's 
atmosphere by solving a complex set of ordinary and partial differential equations. 
It is highly desirable that the numerical methods used for solving those equations produce 
relatively small errors, so that the behavior of an AGCM  reflects the inherent characteristics 
of the continuous model formulation that describes the model developers' understanding of 
the underlying physical and chemical processes \citep[see, e.g.][]{Beljaars:1991, beljaars_et_al:2004, beljaars:2018}. 
However, various studies have shown examples where temporal discretization methods in AGCMs, 
especially those used in the parameterization of unresolved processes or 
the coupling between processes, can produce large errors that significantly affect 
key features of the numerical results 
\citep[e.g.,][]{wan:2013,Gettelman_et_al:2015_MG2_part2,Beljaars_et_al:2017,Donahue_Caldwell:2018,Zhang_et_al:2018,Barrett_et_al:2019}. 
These results are not surprising, given the relatively short timescales 
associated with parameterized processes such as clouds and turbulence and 
the relatively long time steps (typically on the order of tens of minutes) 
used by current global atmospheric GCMs. 

This study attempts to take a first step towards assessing and addressing 
time integration issues associated with physics parameterizations in the 
atmospheric component of the U.S. Department of Energy's Energy Exascale Earth System Model version 1, 
hereafter referred to as EAMv1 \citep{Rasch_et_al:2019,Xie_et_al:2018}. 
The study contains two parts: 
\begin{itemize}
\item 
First, the relative importance of time integration errors 
in present-day climate simulations is assessed for EAMv1.
This is done by using an intuitive and practical metric, 
namely the magnitude of changes in the model's long-term climate 
resulting from a substantial (in our case a factor of 6) reduction of 
the time step sizes used in all major components of the model 
(e.g., resolved dynamics, parameterized radiation, stratiform clouds, 
deep convection, and the numerical coupling of various processes). 
As we show in Section~\ref{sec:quantification}, 
the consequent changes in EAMv1's 10-year climate statistics 
lead to a notable and unsurprising degradation in agreement between 
the simulations and observations because time integration errors which were previously 
compensated by parameter tuning are no longer present, 
and no retuning was performed in this study.
\item 
In the second part, a series of sensitivity experiments is conducted and analyzed 
to identify which components of EAM are responsible for the changes in 
cloud fraction and cloud radiative effects. 
The purpose is to provide clues for future studies that investigate the 
root causes of the sensitivities.
\end{itemize}

The rest of this paper proceeds as follows. 
Section~\ref{sec:model} provides
an overview of the EAM model, introduces the time step sizes used by its main components, 
and briefly describes the numerical methods used for process coupling. 
The common setup of the present-day climate simulations and 
the methods used for assessing statistical significance of the sensitivities
are also described. 
Section~\ref{sec:quantification} presents the impact of a proportional, 
factor-of-6 step size reduction in all major components of EAMv1.
Section~\ref{sec:attribution} presents results from additional numerical experiments to attribute the 
time step sensitivities in cloud fraction and cloud radiative effects presented in Section~\ref{sec:quantification}.
The conclusions are drawn in Section~\ref{sec:conclusions}.

\section{Model and simulation overview}
\label{sec:model}
\subsection{EAMv1}
\label{sec:EAMv1}

EAMv1 is a global hydrostatic AGCM.
The dynamical core solves the so-called primitive equations using a continuous Galerkin spectral-element method for horizontal discretization on a cubed-sphere mesh \citep{Dennis_et_al:2012, Taylor_Fournier:2010}.
The vertical discretization uses a semi-Lagrangian approach in a pressure-based terrain-following coordinate \citep{Lin:2004}.
Main components of the parameterizations suite include 
the solar and terrestrial radiation \citep{Mlawer_et_al:1997,Iacono_et_al:2008},
deep convection \citep{Zhang_McFarlane:1995,Richter_Rasch:2008,Neale_et_al:2008},
turbulence and shallow convection \citep{Golaz_et_al:2002, Larson_et_al:2002, Larson_and_Golaz:2005_MWR,Bogenschutz_et_al:2013}, 
stratiform cloud microphysics \citep{Morrison:2008,Gettelman_et_al:2015_MG2_part1,Gettelman_et_al:2015_MG2_part2,Wang_et_al:2014}, 
aerosol lifecycle and aerosol-cloud interactions \citep{Liu_et_al:2016, Wang_et_al:2019}, 
and land surface processes \citep{Oleson_et_al:2013}.

The so-called low-resolution (or standard) configuration of EAM uses a horizontal grid-spacing of approximately 100~km.
The vertical grid consists of 72 layers covering an altitude range from the Earth's surface to 0.1~hPa (64~km), with layer thicknesses ranging from 20--100~m near the surface to 
about 600~m in the free troposphere up to the lower stratosphere. 
This 1-degree configuration is used as one of the workhorses both 
for model development and for multi-decade simulations targeted at scientific investigations.
A more detailed description of EAMv1 can be found in \citet{Rasch_et_al:2019} and \citet{Xie_et_al:2018}.  

Various time integration methods and time step sizes are used by different parts
(hereafter referred to as components) of EAMv1. 
These are mostly explicit or implicit methods using fixed step sizes. 
For example, in the dynamical core, the temperature, horizontal winds, 
and surface pressure equations are integrated in time
using an explicit five-stage third-order Runge-Kutta method 
\citep{Kinnmark_and_Gray:1984,Guerra_and_Ullrich_2016,Lauritzen_et_al:2018}.
The horizontal tracer advection 
uses a three-stage second-order
strong-stability-preserving Runge-Kutta method 
\citep{Spiteri_and_Ruuth:2002,guba:2014,Lauritzen_et_al:2018}.
A small fraction of the parameterizations, 
e.g., sedimentation of rain and snow and turbulent mixing of aerosols, 
are subcycled using dynamically determined step sizes.

%---- FIGURE ---
\begin{figure}[htbp]
\centering
\includegraphics[width=0.4\textwidth]{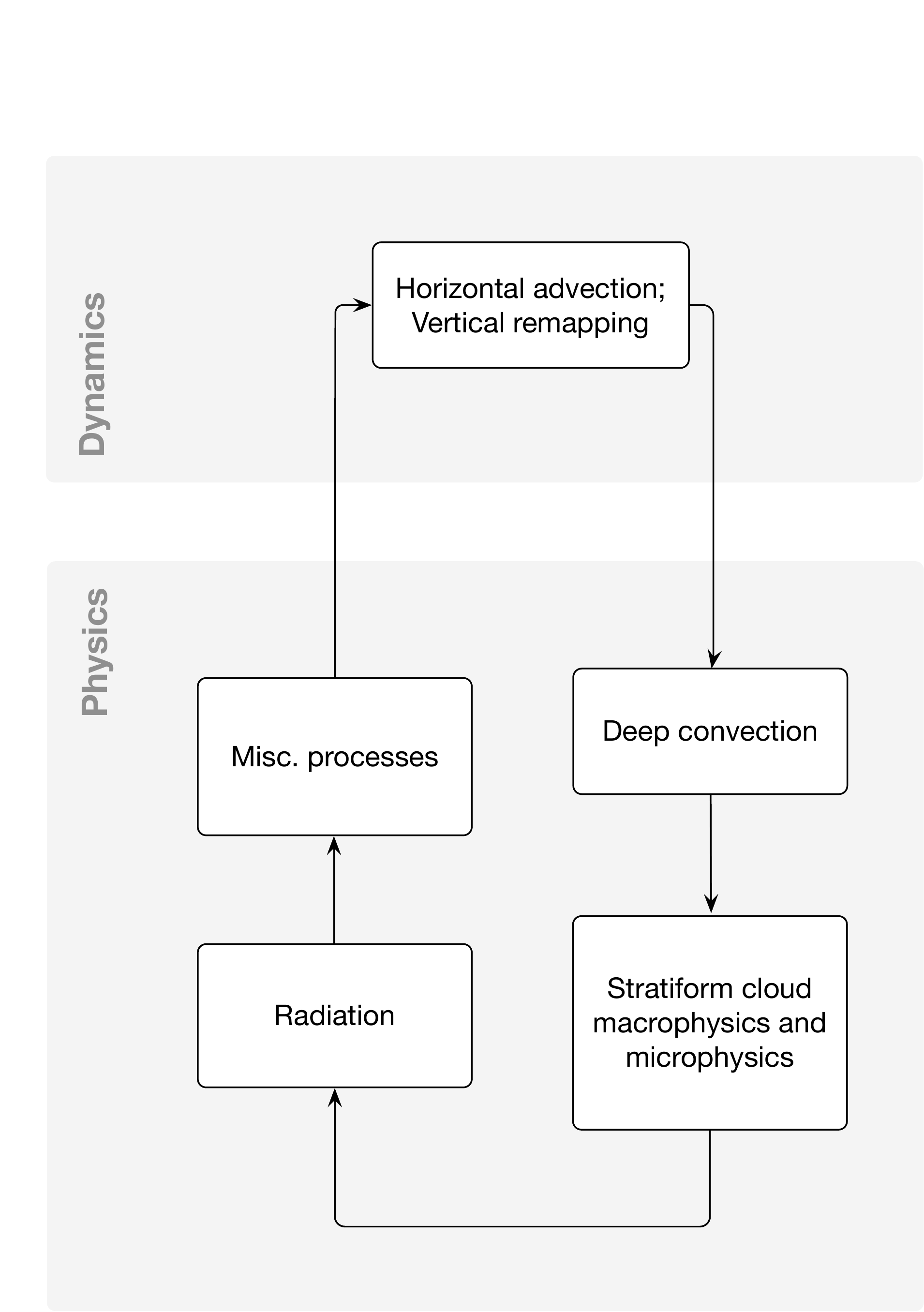}
\caption{
A simplified schematic showing the  
sequence of calculations in EAMv1. Each box is viewed as 
a coarse-grained component (which might contain 
subcomponents corresponding to different atmospheric processes).
Time step sizes used by these coarse-grained components
and their coupling are described in Section~\ref{sec:EAMv1}
and Figure~\ref{fig:schematic_v1_CTRL_and_All_Shorter}a. 
}\label{fig:ordering}
\end{figure}
%------------

%---- FIGURE ---
\begin{figure*}[htbp]
\centering
\includegraphics[width=0.84\textwidth]{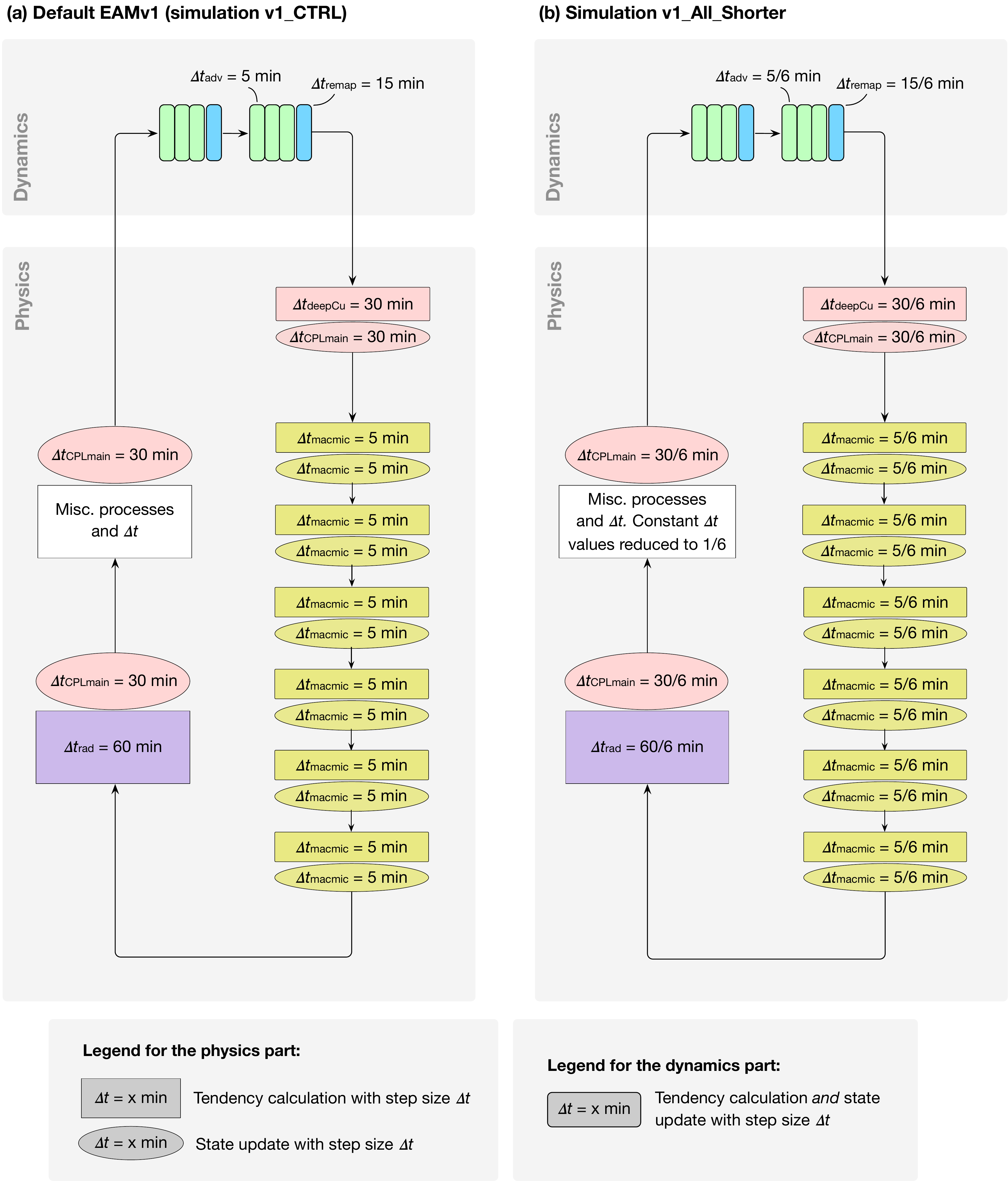}
\caption{
(a) Time step sizes used by the default EAMv1 
at 100~km resolution, corresponding to simulation v1\_CTRL in this paper.
Different colors indicate different step sizes.
Shapes filled with the same color use the same step size.
Further details can be found in Section~\ref{sec:EAMv1}.
(b) Similar to (a) but for the simulation v1\_All\_Shorter 
(cf. Table~\ref{tab:exps} and Table~\ref{tab:exps_namelist}).
\label{fig:schematic_v1_CTRL_and_All_Shorter}
}
\end{figure*}
%------------

These various model components are connected together in 
a sophisticated manner involving 
multiple layers of subcycling and different splitting/coupling methods.
Here we only describe the aspects of process coupling in EAMv1 that 
are investigated in this study. 
Correspondingly, we present in Figure~\ref{fig:ordering} 
a simplified schematic of the process ordering in EAMv1. 
Each box is viewed as a coarse-grained model component
which might contain subcomponents
corresponding to different atmospheric processes.

The primary method used for coupling the components shown in 
Figure~\ref{fig:ordering} is a method we refer to as isolated sequential splitting.
In this method, a model component takes as input the 
atmospheric state variables 
(e.g., winds, temperature, pressure, and tracer concentrations) 
that have already been updated by a preceding component.
Tendencies caused by the current component are calculated
by considering the current component in isolation.
The tendencies are then used to update the atmospheric state before 
passing it to the next component.
(In Figure~\ref{fig:schematic_v1_CTRL_and_All_Shorter} and additional schematics presented later
in the paper, tendency calculations in the physics part are depicted 
by rectangular boxes with sharp corners while the update of 
model state is shown by oval shapes.)
This process splitting/coupling method is 
referred to as ``time splitting''
in \citet{williamson:2002} and in \citet{Lauritzen_et_al:2018}, 
``sequential-update splitting" in \citet{Donahue_Caldwell:2018},
and is often referred to as 
``operator splitting'' in wider numerical modeling communities 
\citep[e.g.,][]{Sportisse:2000}.  
Here,
we use the notation $\Delta t_{\rm CPLmain}$ to denote the step size of the 
splitting/coupling applied 
to the components (boxes) shown in Figure~\ref{fig:ordering}.
In the default 1-degree configuration of EAMv1,
$\Delta t_{\rm CPLmain}$~=~30~min.
The step sizes used in the various components are described below
(see also Figure~\ref{fig:schematic_v1_CTRL_and_All_Shorter}a):
\begin{itemize}

    \item Within the resolved dynamics, the vertical discretization (remapping) uses time steps of $\Delta t_{\rm remap}$= 15~min, each of which is further divided into 3 substeps of $\Delta t_{\rm adv}$= 5~min for the horizontal advection of temperature, momentum, and tracers. These step sizes can be chosen separately as long as $\Delta t_{\rm remap}$ is a multiple of $\Delta t_{\rm adv}$ and $\Delta t_{\rm CPLmain}$ is a multiple of $\Delta t_{\rm remap}$.

    \item Deep convection uses $\Delta t_{\rm deepCu}$= 30~min; this is tied to 
    (i.e., has to be the same as) $\Delta t_{\rm CPLmain}$.  
    
    \item The parameterizations of stratiform and shallow cumulus clouds include two elements: 
(1) a treatment of turbulence and shallow convection using a parameterization named Cloud Layers Unified By Binormals 
\citep[CLUBB,][]{Golaz_et_al:2002, Larson_et_al:2002,Larson_and_Golaz:2005_MWR}, which we refer to 
for brevity as cloud macrophysics in this paper, and 
(2) a treatment for aerosol activation (i.e., the formation of cloud liquid and ice particles) 
and the further evolution of cloud condensate, which we refer to as cloud microphysics. 
These two elements are subcycled together using time steps of 
$\Delta t_{\rm macmic}$= 5~min following \citet{Gettelman_et_al:2015_MG2_part2}. 
CLUBB diagnoses cloud fraction and 
effectively does the large-scale condensation calculation
using its predicted sub-grid probability distribution functions of 
heat, water, and vertical velocity. This means the condensation 
and cloud fraction calculations are done at intervals of $\Delta t_{\rm macmic}$= 5~min.
Within the cloud microphysics parameterization, the sedimentation of hydrometeors 
uses adaptive substepping but the other processes, including for example autoconversion, 
accretion, and self-collection of rain drops etc., 
are calculated using the forward Euler method method with a fixed step size of $\Delta t_{\rm macmic}$. 
Further details about time stepping in the cloud microphysics parameterization 
can be found in Section~2 of \citet{Santos_2020_JAMES_MG2_timescales}. 

To facilitate discussions later in this paper, we use the notation $\Delta t_{\rm CPLmacmic}$ 
to denote the step size used for coupling the collectively subcycled cloud macro- and microphysics 
with the rest of EAM. The default EAMv1 has $\Delta t_{\rm CPLmacmic} \equiv \Delta t_{\rm CPLmain}$ 
while an alternative is discussed in Section~\ref{sec:further_attribution}.  
We note that CLUBB can be further subcycled with respect to $\Delta t_{\rm macmic}$, 
but that is not done in either the default EAMv1 or in any of the simulations presented 
in this paper.
    
    \item Heating/cooling rates resulting from shortwave (SW) and longwave (LW) radiation are calculated every hour, i.e., $\Delta t_{\rm rad}$= 60~min. This means radiation is supercycled with respect to all the other parameterizations as well as the resolved dynamics. During every other time step of $\Delta t_{\rm CPLmain}$=~30~min when the radiation parameterization is not exercised, the tendencies saved from the previous 30~min are used to update the atmospheric state.
    \item Miscellaneous other atmospheric processes, e.g., gravity wave drag and the sedimentation, dry deposition, and microphysics of aerosols, are coupled with each other and with the processes listed above at time intervals tied to $\Delta t_{\rm CPLmain}$. 
    The coupling to land surface happens at intervals of $\Delta t_{\rm CPLmain}$ by default; it can be changed to longer multiples of $\Delta t_{\rm CPLmain}$ but that again is not explored in this study.
\end{itemize} 

These various step sizes are schematically depicted in Figure~\ref{fig:schematic_v1_CTRL_and_All_Shorter}a.
Their relationships in the default EAMv1 can be summarized as follows:
\begin{eqnarray} \label{eq:dt_ratio}
\Delta t_{\rm CPLmain} &=& 2 \Delta t_{\rm remap} = 6 \Delta t_{\rm adv}  \label{eq:dt_dyn}\\
\Delta t_{\rm CPLmain} &\equiv&  \Delta t_{\rm CPLmacmic} = 6 \Delta t_{\rm macmic} \label{eq:dt_macmic}\\
\Delta t_{\rm CPLmain} &\equiv& \Delta t_{\rm deepCu} \label{eq:dt_misc}\\
\Delta t_{\rm CPLmain} &=& 0.5 \Delta t_{\rm rad} \,. \label{eq:dt_rad}
\end{eqnarray} 
The equivalent sign ($\equiv$) indicates step sizes that are tied together in the default EAMv1.

In terms of the coupling among the coarse-grained components shown in Figure~\ref{fig:ordering},
we are currently aware of three instances in which the model state and its tendencies 
are both passed to subsequently calculated components. These instances are: 
\begin{itemize}
\item For the coupling between the parameterized physics and the resolved dynamics,
tendencies of temperature and momentum caused by the 
entire parameterization suite are provided to the dynamical core. 
These are used to update the 
 state variables 
 before each vertical remapping step $\Delta t_{\rm remap}$.
This method of physics-dynamics coupling is depicted in 
Figure~2b of \citet{Zhang_et_al:2018} and also discussed 
in \citet{Lauritzen_and_Williamson_2019_JAMES_energy_error}.
\item Sensible heat fluxes and moisture fluxes at the Earth's surface are calculated in 
the ``Misc. processes'' box in Figure~\ref{fig:ordering}. The fluxes 
are not immediately applied to update the atmospheric state; rather, 
they are passed into the stratiform cloud macro/microphysics subcycles 
and used as boundary conditions for CLUBB.
\item Deep convection is assumed to detrain a certain amount of cloud liquid,
causing a source of stratiform cloud condensate.
The detrainment-induced tendency of stratiform cloud liquid mass concentration 
is not applied within or immediately after the deep convection parameterization but  
passed into the stratiform cloud macro/microphysics subcycles. 
After CLUBB has operated, detrainment-induced cloud mass tendency is partitioned into
liquid and ice phases using the current temperature values; 
temperature tendency corresponding to the effective phase change is diagnosed;
cloud droplet and crystal number tendencies
are derived from the partitioned mass tendencies using assumed 
cloud particle sizes. 
These tendencies of cloud liquid and ice as well as temperature
are used to update the model state variables
before the state variables are provided to the aerosol activation and 
cloud microphysics parameterization.
\end{itemize}
All three cases described above involve passing tendencies of 
some processes (that are calculated with longer step sizes) 
to subsequent processes that are subcycled (i.e., use shorter step sizes).
The spirit of this method resembles the ``sequential splitting'' method
advocated in \citet{beljaars_et_al:2004} and \citet{beljaars:2018}
as well as the ``sequential-tendency splitting'' method defined in \citet{Donahue_Caldwell:2018}.
The method leads to a tighter coupling as the subcycled
processes ``feel'' the influence of the preceding processes
and respond at the shorter intervals; 
this tighter coupling is the motivation for the ``v1\_Dribble'' simulation 
described in Section~\ref{sec:cpl}. 
On the other hand, the processes causing the tendencies respond to 
the subcycled processes only at longer intervals;  
the temporal truncation errors associated with these longer time steps
can be manifested in those tendencies
and hence trigger responses in the subcycled processes.

\subsection{EAMv0}
\label{sec:EAMv0}

To provide context and serve as a reference for the evaluation of time step sensitivity in EAMv1, 
we also present one simulation using EAMv0, i.e., EAMv1's most recent predecessor.
EAMv0 uses the same dynamical core and large-scale transport algorithms as in v1, 
but the vertical grid has only 30 layers. Many of the parameterizations differ from EAMv1.
The parameterization of turbulence and shallow convection follows \citet{Park_Bretherton:2009},
the cloud macrophysics parameterization follows \citet{park:2014},
and the cloud microphysics parameterization is described in \citet{Morrison_Gettelman:2008}.
The time integration methods and step sizes are very similar to those in EAMv1, 
except that the cloud macro- and microphysics parameterizations 
are not subcycled (i.e., they use a 30~min step size).

%---------- TABLE ------------
\begin{sidewaystable}[htbp]
\centering
\caption{List of climate simulations conducted in this study. 
The numbers given in the main part of the table are the ratio of each step size 
(or $\Delta t_{\rm DeepCu}/\tau$) relative to its value in the
default 1-degree model.
The meaning and default values of the various step sizes 
are explained in Section~\ref{sec:model}.
Here $\tau$ refers to the prescribed (fixed) timescale in the deep convection parameterization 
for releasing the convective available potential energy (CAPE), 
the default value of which is 3600~s. 
The namelist variables in EAM used to 
configure these simulations are listed in Table~\ref{tab:exps_namelist}.
Schematics of the EAMv1 simulations are shown in 
Figures~\ref{fig:ordering}, 
\ref{fig:schematic_v1_CTRL_and_All_Shorter}, 
\ref{fig:schematic_macmic_vs_other}, 
\ref{fig:schematic_CPL+DeepCu}, and
\ref{fig:schematic_dribble}.
}
\label{tab:exps}
\begin{adjustbox}{max width=0.95\textwidth}
\begin{tabular}{clll| cc cc ccc |c}
\tophline
\multirow{2}{*}{Group} &
\multirow{2}{*}{Simulation name} &
\multirow{2}{*}{Description} &
\multirow{2}{*}{Schematic} &
\multicolumn{7}{c|}{Ratio of time step size relative to default} & 
\multirow{2}{3cm}{Ratio of $\Delta t_{\rm deepCu}/\tau$ relative to default}\\\cline{5-11}
&&&& $\Delta t_{\rm remap}$ & $\Delta t_{\rm adv}$  & 
  $\Delta t_{\rm CPLmain}$ &
  $\Delta t_{\rm deepCu}$ & 
  $\Delta t_{\rm CPLmacmic}$ &
  $\Delta t_{\rm macmic}$ & 
  $\Delta t_{\rm rad}$ & 
\\\middlehline
0   & v0\_CTRL                          & Sect.~\ref{sec:EAMv0}         & -                               & 1   & 1   & 1   & 1   & 1   & 1   & 1   & 1\\
\middlehline
I   & v1\_CTRL                          & Sect.~\ref{sec:EAMv1}         & Fig.~\ref{fig:schematic_v1_CTRL_and_All_Shorter}a & 1   & 1   & 1   & 1   & 1   & 1   & 1   & 1\\
I   & v1\_All\_Shorter                  & Sect.~\ref{sec:exps}          & Fig.~\ref{fig:schematic_v1_CTRL_and_All_Shorter}b     & 1/6 & 1/6 & 1/6 & 1/6 & 1/6 & 1/6 & 1/6 & 1/6\\
\middlehline
II  & v1\_MacMic\_Shorter               & Sect.~\ref{sec:StCld_vs_rest} & Fig.~\ref{fig:schematic_macmic_vs_other}a     & 1   & 1   & 1   & 1   & 1   & 1/6 & 1   & 1\\
II  & v1\_All\_Except\_MacMic\_Shorter  & Sect.~\ref{sec:StCld_vs_rest} & Fig.~\ref{fig:schematic_macmic_vs_other}b     & 1/6 & 1/6 & 1/6 & 1/6 & 1/6 & 1   & 1/6 & 1/6 \\ 
\middlehline
III & v1\_CPL+DeepCu\_Shorter           & Sect.~\ref{sec:dyn_rad}       & Fig.~\ref{fig:schematic_CPL+DeepCu}     & 1/3 & 1   & 1/6 & 1/6 & 1/6 & 1   & 1   & 1/6\\ 
III & v1\_Dribble                       & Sect.~\ref{sec:cpl}           & Fig.~\ref{fig:schematic_dribble}& 1   & 1   & 1   & 1   & 1/6 & 1   & 1   & 1\\ 
\middlehline
IV & v1\_CPL+DeepCu+Tau\_Shorter       & App.~\ref{sec:dt_tau}       & Fig.~\ref{fig:schematic_CPL+DeepCu}     & 1/3 & 1   & 1/6 & 1/6 & 1/6 & 1   & 1   & 1 \\ 
\bottomhline 
\end{tabular}
\end{adjustbox}
\end{sidewaystable}
%----------------------

\subsection{Present-day climate simulations}
\label{sec:exps}

A series of 10-year simulations were conducted 
using an experimental setup commonly exercised in the development and evaluation 
of EAM and its predecessors.
The model was configured to simulate recent climatological conditions by 
using values of the Earth's orbital conditions, 
aerosol emissions and greenhouse gas concentrations, land use, 
and sea surface temperatures and sea ice coverage 
characteristics of the recent past (around year 2000).
The sea surface temperature and sea ice cover were prescribed using monthly 
climatological values that repeated each year. 
Prognostic equations were integrated in time to produce evolving descriptions 
of the atmosphere and land states. 
The simulations used initial conditions written out by a previously performed multiyear simulation. 
Some of the model configurations used in our sensitivity experiments 
produced climate statistics that differed substantially from the default configuration; 
therefore, to avoid characterizing the initial adjustment phase, 
a 4-month spin-up was performed and neglected in each simulation 
while the 10 subsequent years were analyzed.

\begin{figure*}[htbp]
\vspace{5mm}
\includegraphics[width=0.98\textwidth]{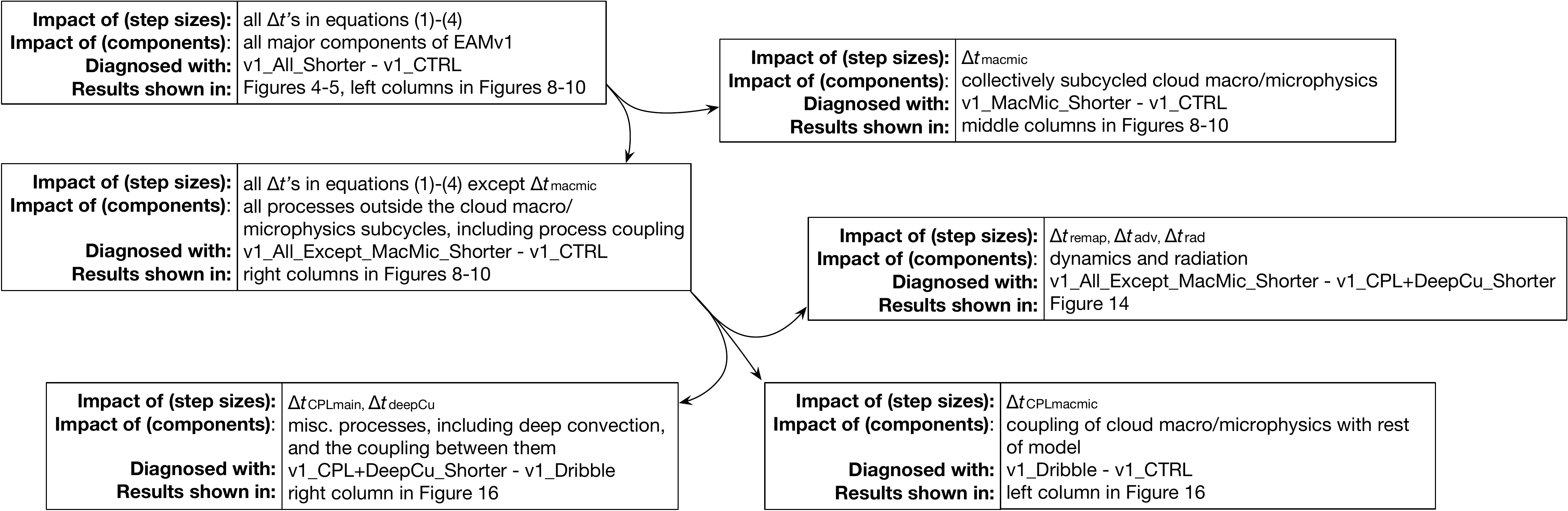}
\caption{A schematic explaining the  
attribution of time step sensitivities.
Time step sizes used in the various simulations are 
summarized in Tables~\ref{tab:exps} and \ref{tab:exps_namelist} and
depicted in
Figures~\ref{fig:schematic_v1_CTRL_and_All_Shorter}, 
\ref{fig:schematic_macmic_vs_other}, 
\ref{fig:schematic_CPL+DeepCu}, and
\ref{fig:schematic_dribble}.
\label{fig:attribution}}
\end{figure*}

Simulations were first conducted with EAMv0 or v1 using their default time steps.
These are labelled ``v0\_CTRL" and ``v1\_CTRL", respectively, in this paper.
In a second v1 simulation called ``v1\_All\_Shorter" 
(cf. Table~\ref{tab:exps} and schematic in Figure~\ref{fig:schematic_v1_CTRL_and_All_Shorter}b), 
the various step sizes listed in Eqs.~\eqref{eq:dt_dyn}--\eqref{eq:dt_rad} were proportionally 
reduced by a factor of 6.
This reduction gives a step size of 5~min for most of the parameterizations 
and the coupling among them, which is significantly shorter than the default 
but is still practically affordable for multiyear sensitivity simulations. 
Results from these three simulations are discussed in Section~\ref{sec:quantification}.
Additional simulations were also conducted with the v1 model
to allow the differences between v1\_All\_Shorter and v1\_CTRL 
to be attributed to specific sets of processes and time stepping algorithms. 
The experimental design is summarized in Tables~\ref{tab:exps} and \ref{tab:exps_namelist}, 
groups II and III. 
The attribution process is summarized in Figure~\ref{fig:attribution}
with the detailed results discussed in Section~\ref{sec:attribution}.

%---------- figure ------------
%----------------------

\subsection{Statistical tests}
\label{sec:t-tests}

The analyses presented in this paper focus primarily on 10-year mean annual averages.
To distinguish signals of time step sensitivity from noise caused by natural variability,
the two-sample $t$-test was applied to pairs of simulations,
with the test statistic constructed using annual averages.
A significance level of 0.05 was chosen to determine whether 
differences between a pair of 10-year averages were statistically significant.
This method of two-sample $t$-test has been used in the diagnostics package 
from the
National Center for Atmospheric Research (NCAR) Atmosphere Model Working Group (AMWG)
who developed predecessors of EAM 
(\url{http://www.cesm.ucar.edu/working_groups/Atmosphere/amwg-diagnostics-package/}) .

Considering that the sample size of 10 is relatively small, 
we also conducted statistical testing using monthly mean model output.
Serial correlation in monthly averages was addressed by
using the paired $t$-test and the effective sample size \citep{Zwiers_and_vonStorch_1995}.
For example, to assess the significance of the differences between 
simulations $A$ and $B$ at a certain geographical location, 
we used the time series of monthly mean $A-B$ (which had 120 data points in the monthly time series)
to construct the test statistic for a one-sample $t$-test.
A significance level of 0.05 was chosen to determine whether 
the mean of the differences was statistically zero, 
taking into account the autocorrelation in the time series.

We processed all the difference plots shown in the paper using both methods. 
The two methods turned out to give rather consistent results overall. 
They disagree only at a small portion
of grid points associated with relatively small climate differences.
The key signatures of time step sensitivity discussed below were considered statistically 
significant by both methods.
We chose to show results from the two-sample test 
here, to be consistent with the AMWG diagnostics package.

\section{Impact of proportional step size reductions in all major processes}
\label{sec:quantification}

The first question we attempt to answer is whether the characteristics of EAMv1's present-day climate are substantially affected by the choices of time step sizes. This is done by comparing simulations
v1\_CTRL and v1\_All\_Shorter. 
To put the magnitude of the differences into context,
we also show some representative results from v0\_CTRL.

%---- FIGURE ---
\begin{figure*}[htbp]
\vspace{-0.1in}
\centering
\includegraphics[width=0.85\textwidth]{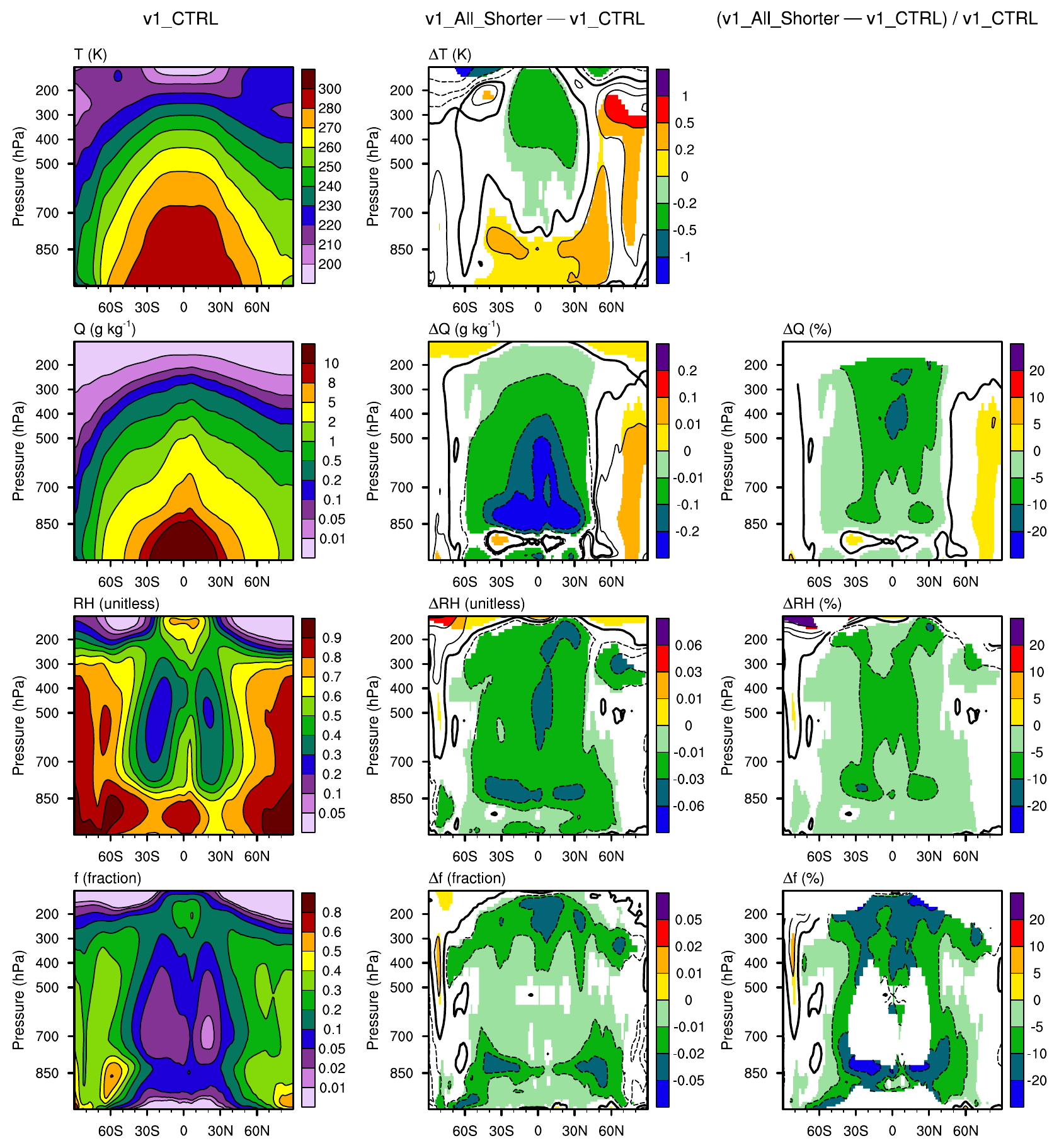}
\caption{
Left column: 10-year mean, zonally averaged air temperature (T), 
specific humidity (Q),
relative humidity (RH), and cloud fraction (f) in simulation v1\_CTRL. 
Middle column: differences between v1\_All\_Shorter and v1\_CTRL.
Right column: relative differences with respect to v1\_CTRL.
Statistically insignificant differences are masked out in white.
The simulation setups are described in Section~\ref{sec:exps}
and also summarized in group I in Tables~\ref{tab:exps} and \ref{tab:exps_namelist}.
Schematics depicting the time integration loop and 
different step sizes can be found in Figure~\ref{fig:schematic_v1_CTRL_and_All_Shorter}.
}
\label{fig:5min_vs_30min_presheight}
\end{figure*}
%--------------

%---- FIGURE ---
\begin{figure*}[htbp]
\vspace{-0.1in}
\centering
\includegraphics[width=0.95\textwidth]{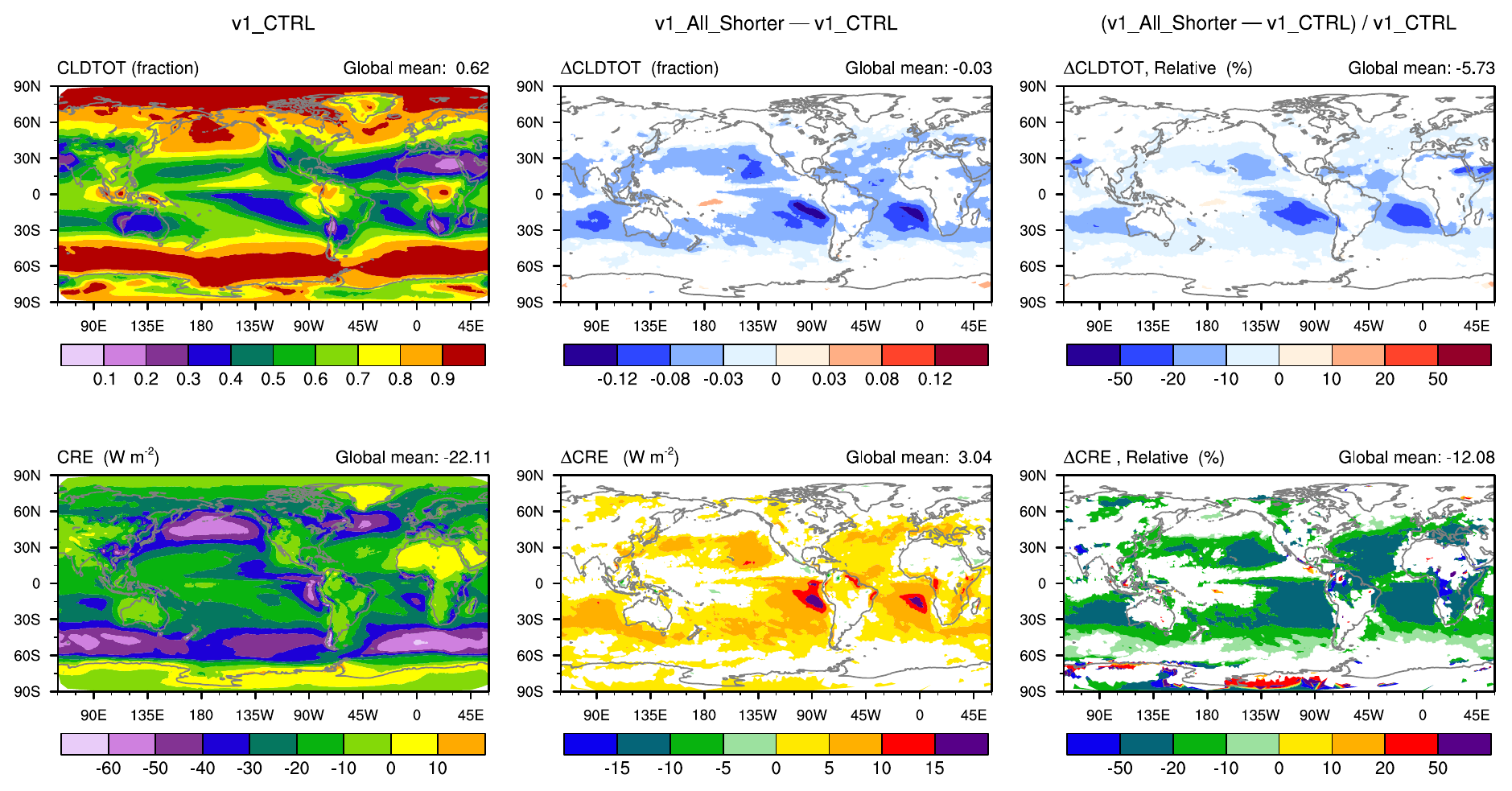}
\caption{
Left column: 10-year mean geographical distribution of
total cloud cover (CLDTOT, upper row) and
total cloud radiative effect (CRE, lower row) in v1\_CTRL. 
Middle column: differences between v1\_All\_Shorter and v1\_CTRL.
Right column: relative differences with respect to v1\_CTRL.
Statistically insignificant differences are masked out in white.
The simulation setups are described in Section~\ref{sec:exps}
and also summarized in group I in Tables~\ref{tab:exps} and \ref{tab:exps_namelist}.
Schematics depicting the time integration loop and 
different step sizes can be found in Figure~\ref{fig:schematic_v1_CTRL_and_All_Shorter}.
}
\label{fig:5min_vs_30min_maps}
\end{figure*}
%--------------

\subsection{Time step sensitivities in EAMv1}

It turns out that the proportional factor-of-6 step size reduction in all major components 
of the v1 model leads to systematic changes in the simulated long-term climate. 
In the middle column of Figure~\ref{fig:5min_vs_30min_presheight}, 
the differences in 10-year-mean zonal averages between v1\_All\_Shorter and v1\_CTRL 
are shown for air temperature (T), specific humidity (Q), 
relative humidity (RH) and cloud fraction (f).
The relative differences normalized by corresponding values in v1\_CTRL
are shown in the right column. (Relative differences in T are not a useful measure
and hence not included.) 
Statistically {\it insignificant} differences are masked out in white.
The figure reveals that the 
step size reduction leads to warming of 
up to 0.5~K in the subtropical and mid-latitude near-surface layers
and cooling of similar magnitudes in the tropical mid- and upper-troposphere
(Figure~\ref{fig:5min_vs_30min_presheight}, second panel in first row).
In the middle and low latitudes, the air dries at most altitudes in the troposphere,
showing typical decreases of 1\% to 10\% 
in both specific and relative humidity
(Figure~\ref{fig:5min_vs_30min_presheight}, second and third rows).
Cloud fraction also decreases (Figure~\ref{fig:5min_vs_30min_presheight}, bottom row); 
the largest changes appear in three regions: 
in the upper troposphere where ice clouds dominate,
in the subtropical lower troposphere where stratocumulus and trade cumulus prevail, 
and in the mid-latitude near-surface layers.

The 10-year mean geographical distributions of total cloud cover and total cloud radiative effect (CRE) are shown in Figure~\ref{fig:5min_vs_30min_maps}. Here the signatures of time step sensitivity appear to be dominated by changes in subtropical marine stratocumulus and trade cumulus clouds.
The largest local changes are on the order of -10\% to -50\% for cloud cover and -20\% to -50\% for CRE.
The global mean CRE weakens by about 3~W~m$^{-2}$, corresponding to a relative change of -12\%.

\subsection{Comparison with observations and EAMv0}
\label{sec:v0}

\begin{figure*}[htbp]
\centering
\includegraphics[width=0.7\textwidth]{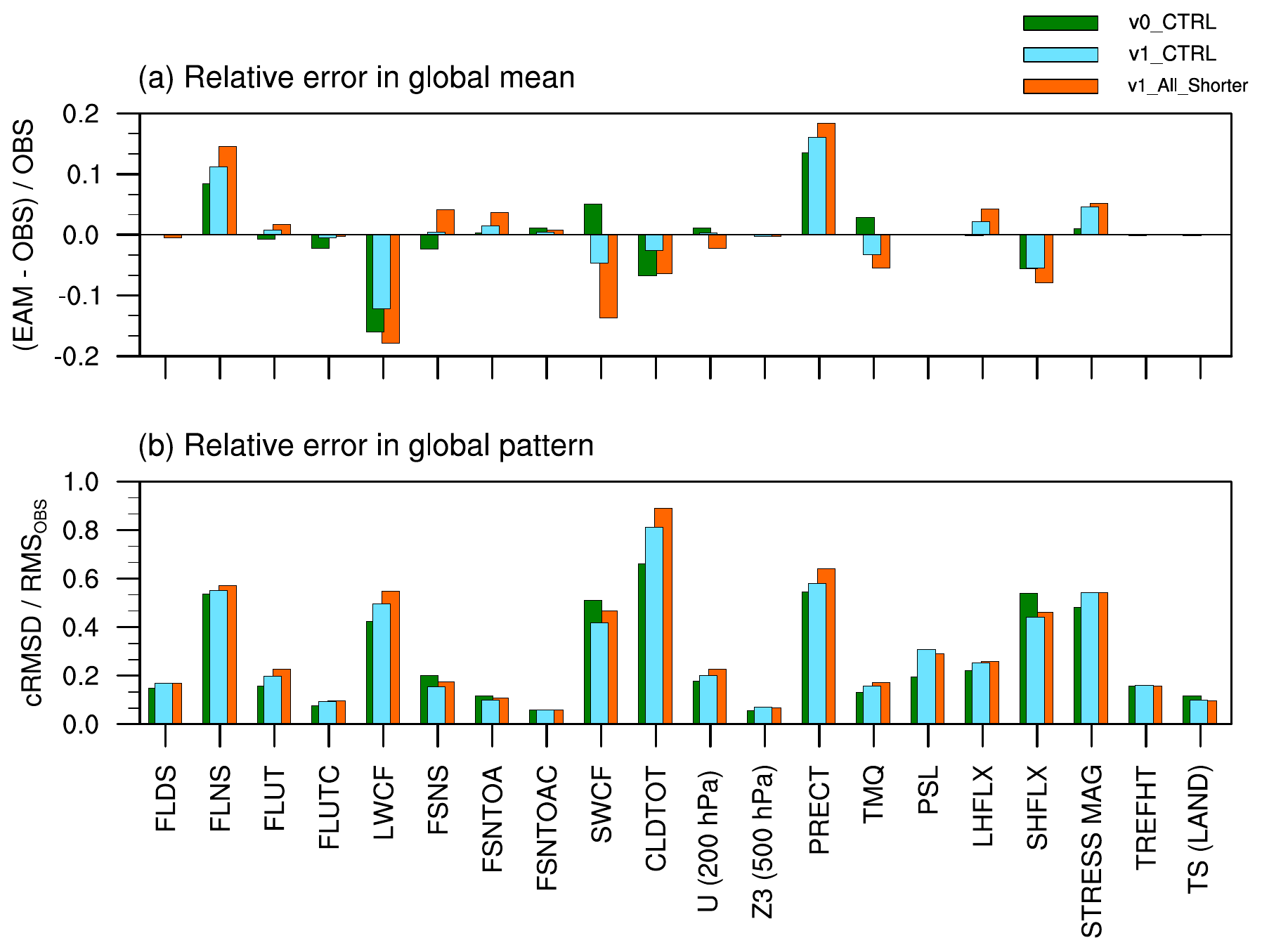}
\caption{
Comparison of 10-year-mean climate simulated by v1\_All\_Shorter, v1\_CTRL and v0\_CTRL
against various reanalyses and satellite products.
The upper panel shows 
relative errors in the simulated global averages.
The lower panel shows the relative error in the simulated geographical distributions, as measured by the centered 
root-mean-square differences (cRMSD) between model results and the observations
normalized by the root-mean-square of the observed global distribution (RMS$_{\rm OBS}$).
The long names of the physical quantities labeled along the x-axis
are given in Table~\ref{tab:obsvar} together with the sources of observational data.
}
\label{fig:error_eam_obs}
\end{figure*}

A recent evaluation of EAMv1 has shown that
the simulated present-day climate is 
cooler and drier than reanalysis in the tropical upper troposphere 
while the CRE in the major marine stratocumulus regions are weaker
compared to satellite products
\citep[cf. Figures~3, 4, and 10 in][]{Rasch_et_al:2019}.
Comparing those results 
with the time step sensitivities shown in
Figures~\ref{fig:5min_vs_30min_presheight} and 
\ref{fig:5min_vs_30min_maps}, one gets the impression that model biases in v1\_All\_Shorter
are likely to be larger than those in v1\_CTRL. 
Here a model bias is defined as a deviation from the real-world observation.
To obtain a comprehensive and yet concise assessment of the impact of time step sizes 
on model fidelity, we follow the spirit of Figure~2 in \citet{Donahue_Caldwell:2018}
and use the collection of reanalyses and satellite products 
listed in Table~\ref{fig:error_eam_obs}
to evaluate the fidelity of v1\_CTRL and v1\_All\_Shorter.
The results are presented in Figure~\ref{fig:error_eam_obs},
where the upper panel shows 
the relative errors in the simulated global averages and 
the lower panel shows the relative errors in global patterns. 
The relative  pattern error is defined as the centered root-mean-square (RMS) 
difference between the simulated and observed patterns, 
normalized by the RMS of the observed pattern. 
A ``pattern" here refers to the annual mean, global, geographical distribution
of a physical quantity.
The model results used in the calculations were 10-year averages.
The observational data was averaged over the years indicated in Table~\ref{fig:error_eam_obs}.
The biases in v0\_CTRL are also included in the figure for comparison.

%-----------TABLE--------
\begin{table*}[htbp]
\caption{List of observational data and EAM's output variables
used for evaluating model biases. 
The observational data were obtained from NCAR's AMWG diagnostics package 
(\url{http://www.cgd.ucar.edu/amp/amwg/diagnostics/plotType.html}). 
TOA stands for Top Of Atmosphere. 
}
\label{tab:obsvar}
\centering
%\begingroup
%\setlength{\heavyrulewidth}{1.5pt}
%\setlength{\abovetopsep}{4pt}
%\setlength{\tabcolsep}{6pt} % Default value: 6pt
%\renewcommand{\arraystretch}{1.2} % Default value: 1
\begin{tabular}{l l l l}
\hline
Physical quantity                       & Source of observation      & EAM output \\
\hline
Surface longwave downwelling flux       & ISCCP (1983--2000)             & \quad FLDS             \\
Surface net longwave flux               & ISCCP (1983--2000)             & \quad FLNS             \\
TOA upward longwave flux                & CERES-EBAF (2000--2010)  & \quad FLUT             \\
TOA clearsky upward longwave flux       & CERES-EBAF (2000--2010)  & \quad FLUTC          \\
TOA longwave cloud forcing              & CERES-EBAF (2000--2010)  & \quad LWCF            \\
Surface net shortwave flux              & ISCCP  (1983--2000)            & \quad FSNS             \\
TOA net shortwave flux                  & CERES-EBAF (2000--2010)  & \quad FSNTOA         \\
TOA clearsky net shortwave flux         & CERES-EBAF (2000--2010)  & \quad FSNTOAC      \\
Shortwave cloud radiative effect        & CERES-EBAF (2000--2010)  & \quad SWCF            \\
Total cloud amount                      & CloudSat (2007--2010)         & \quad CLDTOT        \\
200 hPa zonal wind                      & JRA25 (1979--2004)              & \quad U            \\
500 hPa geopotential height             & JRA25 (1979--2004)              & \quad Z3          \\
Precipitation rate                      & GPCP (1979--2009)               & \quad PRECT           \\
Total precipitable water                & NVAP  (1988--1999)              & \quad TMQ         \\                                                        
\                                                        
Sea level pressure                      & ERAI (1989--2005)                 & \quad PSL                 \\
Surface latent heat flux                & JRA25 (1979--2004)               & \quad LHFLX             \\
Surface sensible heat flux              & JRA25 (1979--2004)              & \quad SHFLX             \\
Surface stress                          & ERS (1992--2000)                  & \quad TAUX, TAUY  \\
2m air temperature                      & LEGATES (1920--1980)         & \quad TREFHT           \\         
Sea level temperature on land           & NCEP (1979--1998)               & \quad TS          \\                                                                                                                       
\hline
 \end{tabular}
% \endgroup
\end{table*}
%%%-------------------

%---- FIGURE ---
%-------------------

Figure~\ref{fig:error_eam_obs} reveals that
model biases in both the global mean (upper panel) and 
the spatial pattern (lower panel) 
are larger in v1\_All\_Shorter for most of the physical
quantities examined here; 
the magnitude of the differences is comparable to the
differences between v1\_CTRL and v0\_CTRL.
For clarification, we note that v1\_CTRL and v0\_CTRL
have different characteristic biases due to the substantial
changes in the parameterizations and vertical resolution.
For example, Figure~\ref{fig:SWCF_v0_v1_vs_obs} shows that 
the shortwave CRE biases in the low latitudes 
are dominated by overestimation in the monsoon regions in v0
and underestimation associated with the marine stratocumulus 
decks and over the warm pool in v1.
If we compare the local differences between 
v1\_All\_Shorter and v1\_CTRL with the local differences 
between v0\_CTRL and v1\_CTRL, then 
the time step caused differences will appear to be 
substantially smaller than the differences caused 
by changes in parameterizations and vertical resolution, 
as should be expected. On the other hand, 
when comparing all three simulations (v1\_All\_Shorter, v1\_CTRL, and v0\_CTRL)
with observations using the metrics shown in Figure~\ref{fig:error_eam_obs}, 
we see that the degradation 
of model fidelity caused by reducing step sizes in v1 
has a magnitude similar to the fidelity improvements from v0\_CTRL to v1\_CTRL.

Given that substantial efforts have been made to tune the default EAMv1,
i.e., to adjust the values of uncertain parameters in the model's equations in order to improve the match between the simulations and observations
\citep[see, e.g.,][]{Xie_et_al:2018,Rasch_et_al:2019},
a degradation of model fidelity associated with shortened time steps
is not surprising.
Assuming the time integration methods used in EAMv1 are mathematically
consistent and convergent, one would expect 
shorter time steps to give {\it numerically} more accurate results. 
The results shown in Figure~\ref{fig:error_eam_obs}
indicate that the default EAMv1 contains sizable time integration errors 
that are compensated by parameter tuning or by other sources of model error.
While the existence of compensating errors is undesirable, 
it is the widely recognized and accepted status quo.  
Reducing time integration errors would sacrifice the immediate results 
and temporarily degrade model fidelity, but it would also provide the opportunity to 
first expose and then address errors from other sources, 
hence could eventually lead to a model that gives correct results for 
correct reasons. 
As a first step towards reducing time-stepping errors in EAMv1, 
the next section identifies the model components that
have caused the differences 
between v1\_All\_Shorter and v1\_CTRL.
While a number of physical quantities are shown 
in Figure~\ref{fig:error_eam_obs},
the analysis in the remainder of the paper focuses on
cloud fraction and CRE. Extension of the analysis to 
additional variables, such as temperature, humidity, precipitation, and winds,
is left to future studies.

\section{Attributing time step sensitivities in cloud fraction and CRE}
\label{sec:attribution}

The primary method used  here for attributing the differences 
between v1\_All\_Shorter and v1\_CTRL is to carry out sensitivity experiments 
in which we vary the step sizes used by different subsets of EAM's components.
These experiments are summarized in 
Tables~\ref{tab:exps} and \ref{tab:exps_namelist} (see groups II and III therein),
Figures~\ref{fig:schematic_macmic_vs_other}, \ref{fig:schematic_CPL+DeepCu} and \ref{fig:schematic_dribble},
and are described in detail
in the following subsections.
An overview of the attribution process is provided in Figure~\ref{fig:attribution}.

%====================

\subsection{Stratiform cloud parameterizations versus the rest of EAMv1}
\label{sec:StCld_vs_rest}

The key signatures in the geographical distribution of total cloud cover and CRE changes 
are seen in the subtropics where marine stratocumulus and trade cumulus 
are the dominant cloud types 
(Figure~\ref{fig:5min_vs_30min_maps}). 
Since these clouds are strongly affected by turbulence, shallow convection,
and cloud microphysics,
it seems natural to link the observed time step sensitivities to the corresponding parameterizations. 
Two hypotheses are explored here:

%---- FIGURE ---
\begin{figure*}[htbp]
\centering
\includegraphics[width=0.84\textwidth]{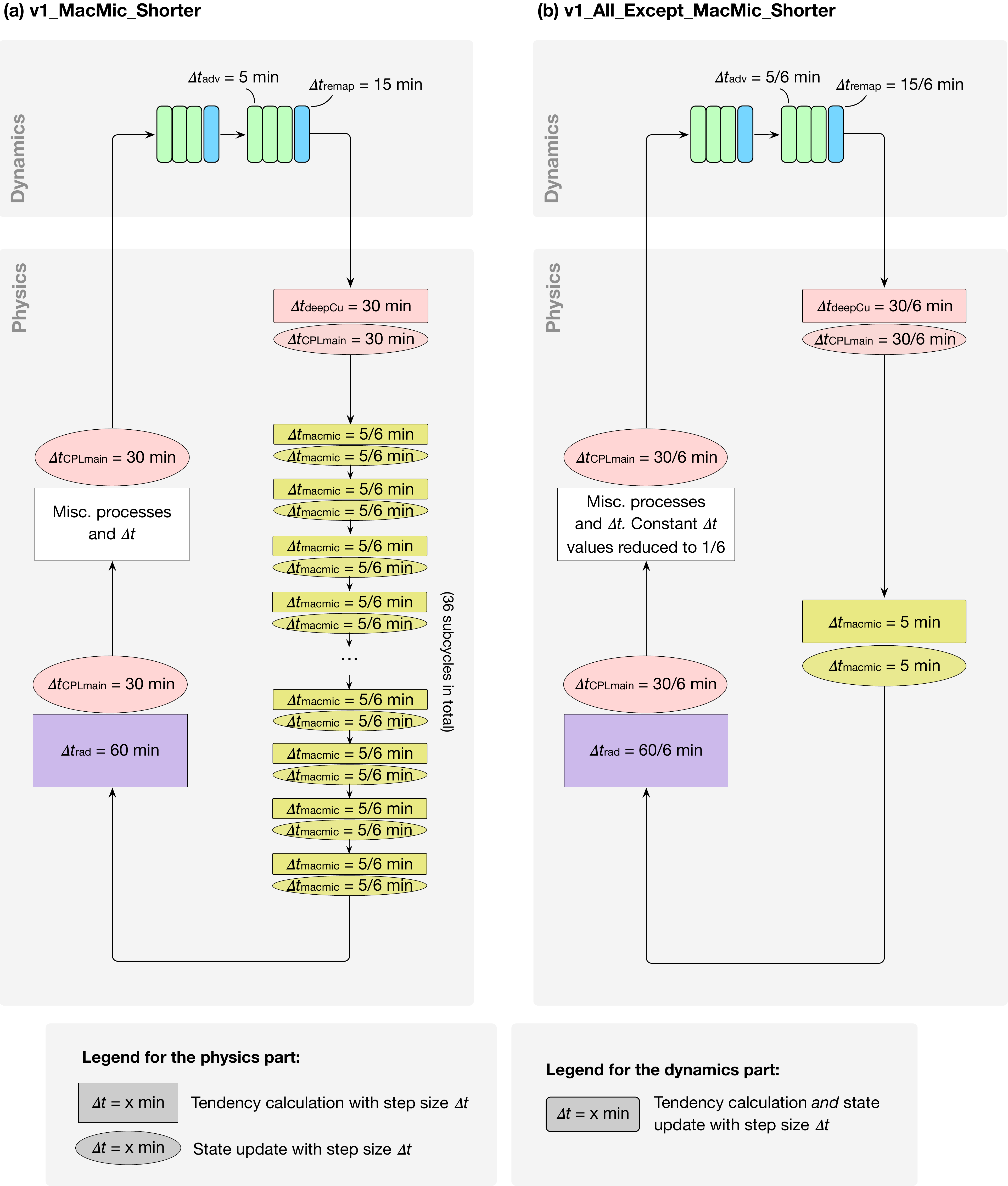}
\caption{
(a) Simulation v1\_MacMic\_Shorter where
the time steps of the collectively subcycled stratiform cloud macro- and microphysics 
are shortened to 1/6 of the default value, i.e., $\Delta t_{\rm macmic}$~=~5/6~min
instead of 5~min.
(b) Simulation v1\_All\_Except\_Macmic\_Shorter
where $\Delta t_{\rm macmic}$ is kept at its default of 5~min while
step sizes for the other parts of EAMv1 were shortened to 1/6.
The color coding follows Figure~\ref{fig:ordering}a.
The simulation setups are summarized in 
Tables~\ref{tab:exps} and \ref{tab:exps_namelist}.
The results are discussed in in Section~\ref{sec:StCld_vs_rest}.
}
\label{fig:schematic_macmic_vs_other}
\end{figure*}

\begin{itemize}

\item {Hypothesis 1:}
The differences in total cloud cover and CRE seen in the subtropics
between v1\_All\_Shorter and v1\_CTRL are caused by 
time integration errors in
the stratiform and shallow cumulus cloud macro- and microphysics parameterizations, 
i.e., CLUBB, aerosol activation, and MG2. 
Turbulence and cloud microphysics are known 
to have relatively short characteristic timescales. 
The 5~min step size ($\Delta t_{\rm macmic}$=~5~min)
used in the default EAMv1 cannot sufficiently resolve those 
short timescales and hence gives numerically inaccurate
results. 

\item {Hypothesis 2:}
The differences in total cloud cover and CRE seen in the subtropics
between v1\_All\_Shorter and v1\_CTRL 
are  caused by time integration errors in parts of EAM 
other than the cloud macro- and microphysics parameterizations 
or in process coupling.
In v1\_All\_Shorter, the reduction of time integration error in those other
components, or their coupling with cloud macro- and microphysics, 
results in a different atmospheric environment 
being provided to CLUBB, and hence triggering 
different responses in shallow cumulus and stratiform clouds.

\end{itemize}
The two sensitivity experiments listed in group II of 
Tables~\ref{tab:exps} and \ref{tab:exps_namelist}
were carried out to test the two hypotheses:
Simulation {\bf v1\_MacMic\_Shorter} (cf. schematic in Figure~\ref{fig:schematic_macmic_vs_other}a)
sets the step size of the collectively subcycled
shallow cumulus and stratiform cloud parameterizations, 
i.e., CLUBB, aerosol activation, and MG2,
to 1/6 of the default value, i.e,
$\Delta t_{\rm macmic}$~=~5/6~min as in v1\_All\_Shorter. 
The rest of EAMv1 used the same time integration strategy 
(and thus the same time step sizes) as in v1\_CTRL.
In other words, within each of the main coupling time step $\Delta t_{\rm CPLmain}$~=~30~min,
instead of 6 invocations of the cloud macro- and microphysics 
parameterizations with 5~min time steps,
there were 36 invocations with 5/6~min time steps.
The differences between results from v1\_MacMic\_Shorter and v1\_CTRL
are attributed to differences in $\Delta t_{\rm macmic}$ which 
controls the step sizes used by CLUBB and MG2 as well as 
the interactions between the processes within each subcycle.

Simulation {\bf v1\_All\_Except\_MacMic\_Shorter} 
(cf. schematic in Figure~\ref{fig:schematic_macmic_vs_other}b) 
has the opposite setup,
i.e., using $\Delta t_{\rm macmic}$=~5~min as in v1\_CTRL,
while the rest of EAMv1 used the much shorter steps employed in v1\_All\_Shorter.
The differences in model climate between v1\_All\_Except\_MacMic\_Shorter and v1\_CTRL
are attributed to reduced step sizes for all model components outside the 
cloud macro- and microphysics subcycles and for the coupling 
(i.e., information exchange) between the subcycles and the other components 
(cf. Eqs.~\eqref{eq:dt_dyn}--\eqref{eq:dt_rad}).

\begin{figure*}[htbp]
\centering
\includegraphics[width=0.8\textwidth]{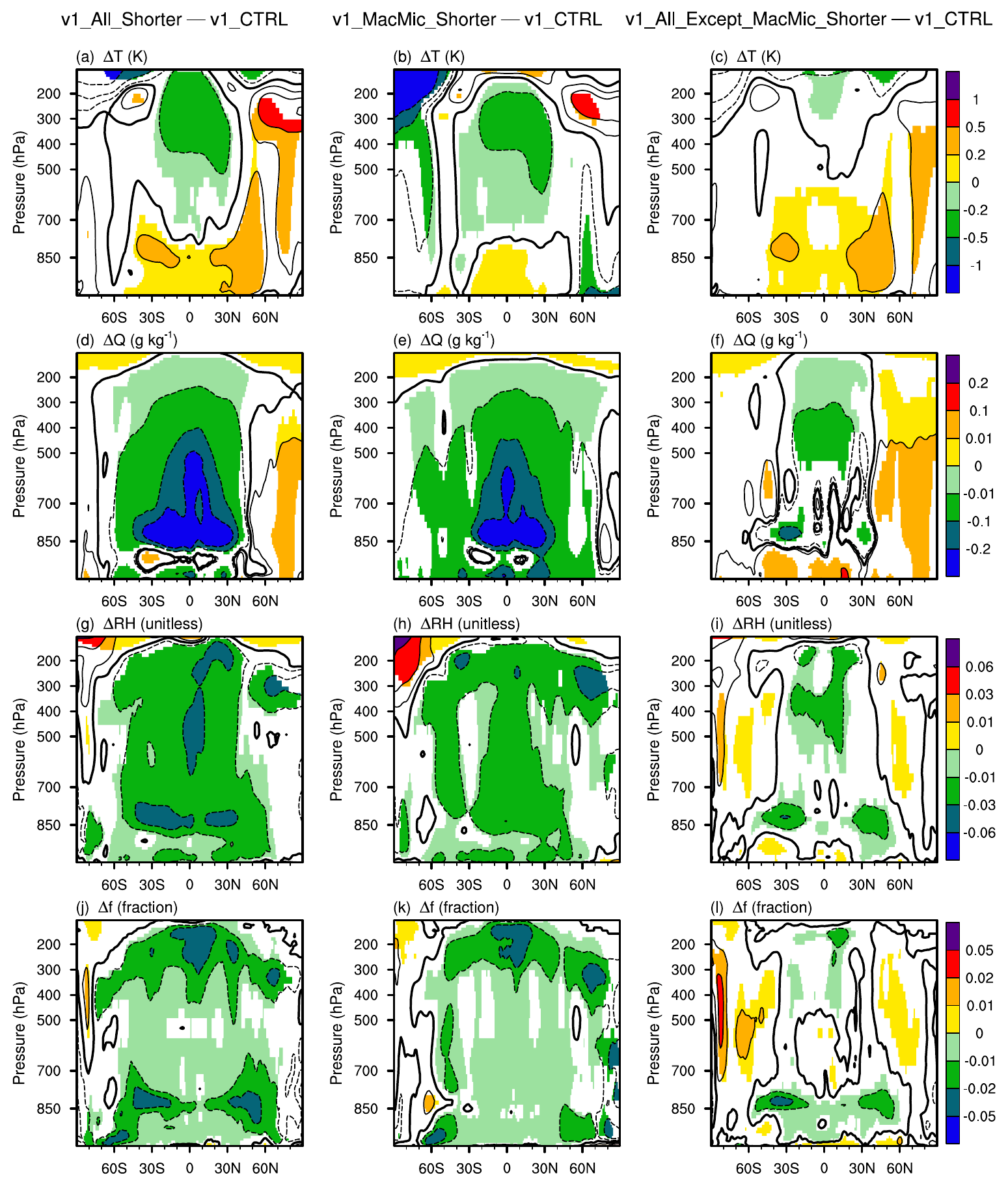}
 \caption{
Differences in 10-year mean, zonally averaged air temperature (T), 
specific humidity (Q),
relative humidity (RH), and cloud fraction (f) between various simulations.
Left column: v1\_All\_Shorter - v1\_CTRL,
revealing the impact of shortening all major time steps listed in Eqs.~\eqref{eq:dt_dyn}-\eqref{eq:dt_rad}; 
middle column: v1\_MacMic\_Shorter - v1\_CTRL,
revealing the impact of shortening time steps for the subcycled cloud macro- and microphysics 
parameterizations; 
right column: v1\_All\_Except\_MacMic\_Shorter - v1\_CTRL,
revealing the impact of shortening step sizes outside the cloud macro- and microphysics subcycles.
Statistically insignificant differences are masked out in white.
The simulation setups are summarized in Tables~\ref{tab:exps} and \ref{tab:exps_namelist}.
Schematics depicting the time integration loop and 
different step sizes can be found in 
Figures~\ref{fig:schematic_v1_CTRL_and_All_Shorter} and \ref{fig:schematic_macmic_vs_other}.
}
\label{fig:macmic_vs_other_presheight}
\end{figure*}

\begin{figure*}[htbp]
\centering
\includegraphics[width=0.95\textwidth]{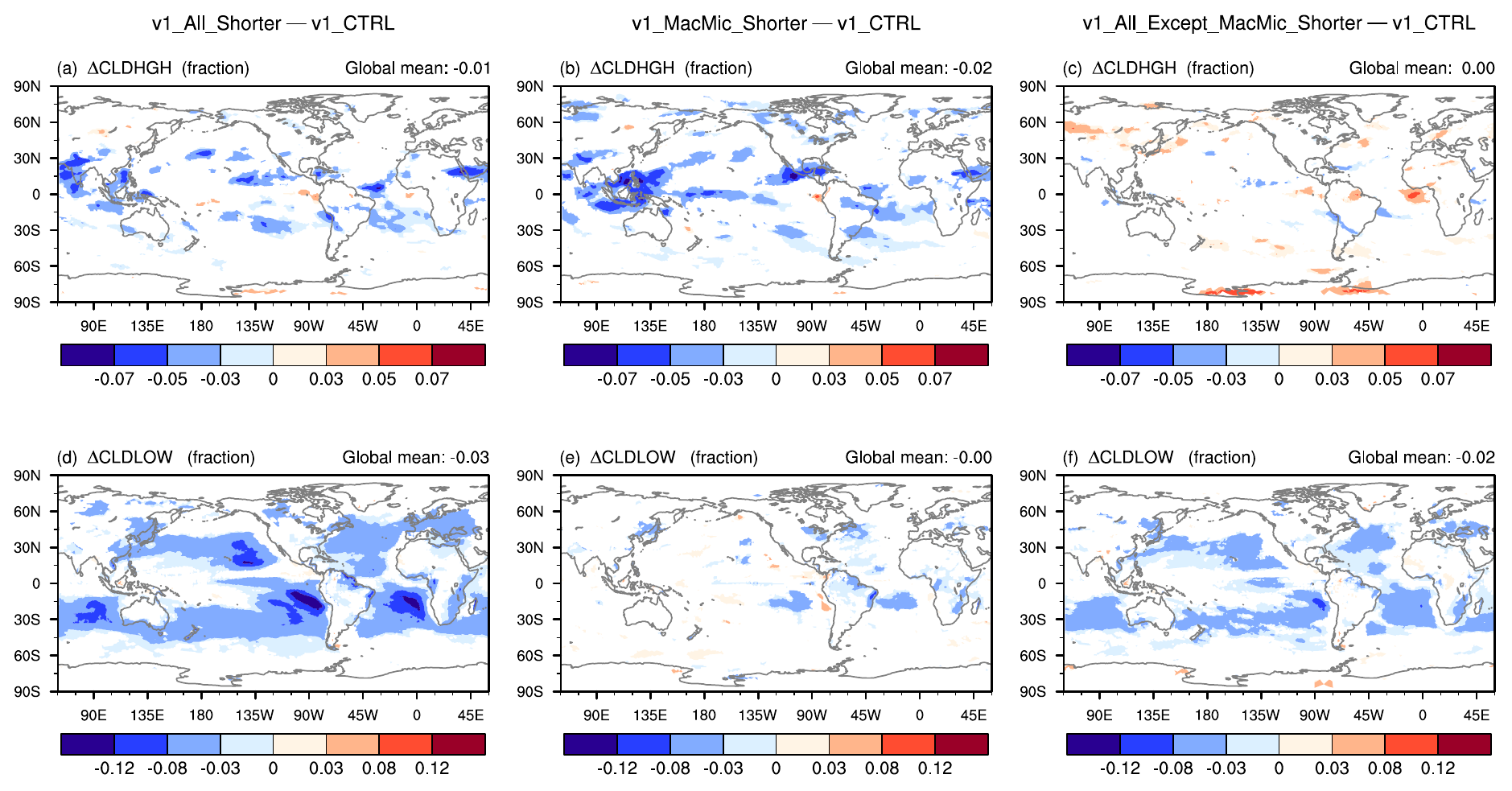}
 \caption{
 Geographical distribution of 10-year mean differences in high-cloud fraction (CLDHGH, upper row)
 and low-cloud fraction (CLDLOW, bottom row). 
 Left column: differences between v1\_All\_Shorter and v1\_CTRL,
 revealing the impact of shortening all major time steps listed in Eqs.~\eqref{eq:dt_dyn}-\eqref{eq:dt_rad}; 
 middle column: differences between v1\_MacMic\_Shorter and v1\_CTRL,
 revealing the impact of shortening time steps for the subcycled cloud macro- and microphysics 
 parameterizations; 
 right column: differences between v1\_All\_Except\_MacMic\_Shorter and v1\_CTRL,
 revealing the impact of shortening step sizes outside the cloud macro- and microphysics subcycles. 
Statistically insignificant results are masked out in white.
The simulation setups are summarized in Tables~\ref{tab:exps} and \ref{tab:exps_namelist}.
Schematics depicting the time integration loop and 
different step sizes can be found in 
Figures~\ref{fig:schematic_v1_CTRL_and_All_Shorter} and \ref{fig:schematic_macmic_vs_other}.
}
\label{fig:macmic_vs_other_maps_f}
\end{figure*}

\begin{figure*}[htbp]
\centering
\includegraphics[width=.95\textwidth]{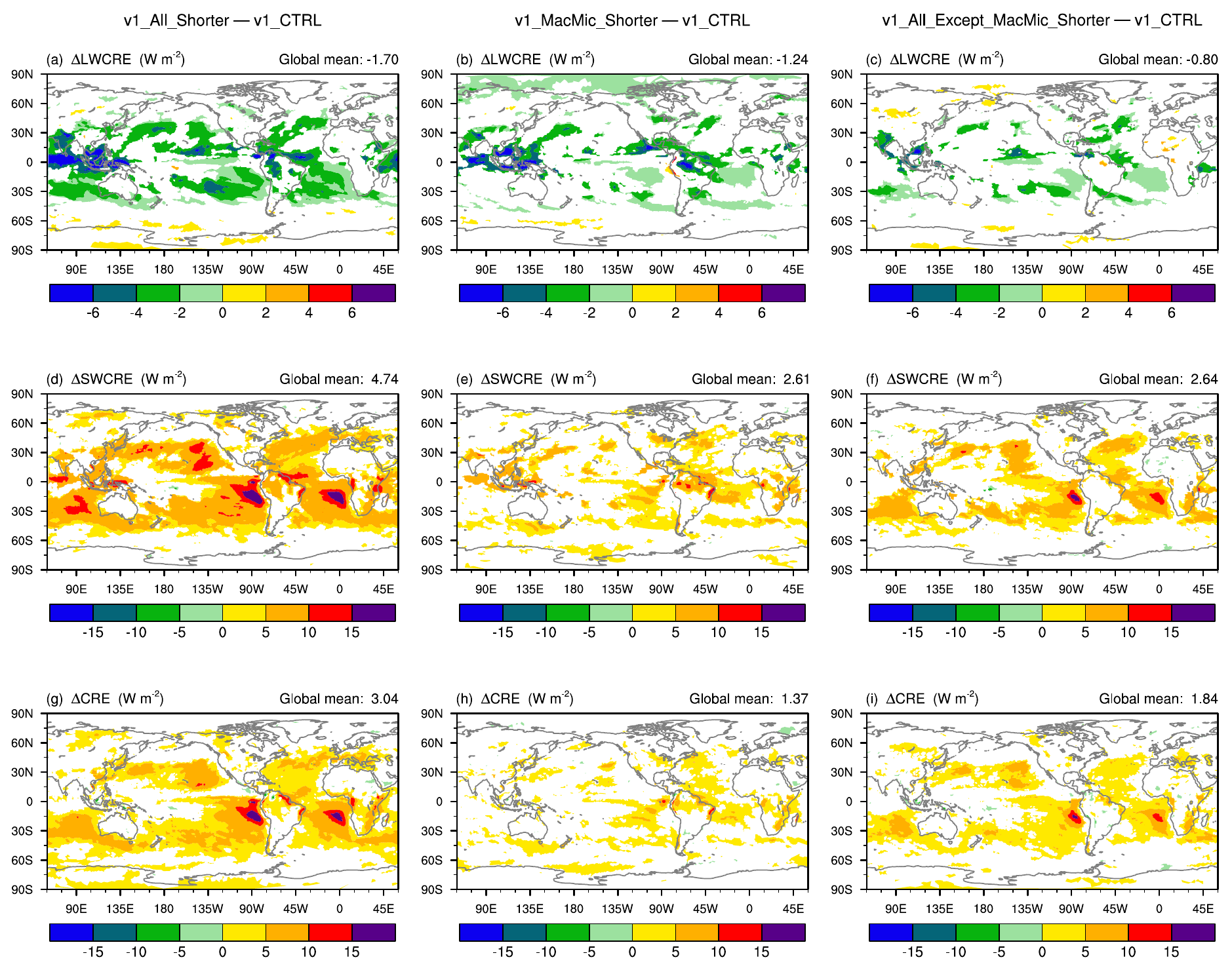}
\caption{
As in Figure~\ref{fig:macmic_vs_other_maps_f}, but showing 
the longwave (LW, top row), shortwave (SW, middle row), and total (bottom row) CRE.
 }
\label{fig:macmic_vs_other_maps_CRE}
\end{figure*}

The 10-year mean difference plots shown in 
Figures~\ref{fig:macmic_vs_other_presheight}--\ref{fig:macmic_vs_other_maps_CRE}
indicate that the changes in long-term climate caused by $\Delta t_{\rm macmic}$ and 
the other step sizes are both non-negligible, but have different signatures.
The zonal mean temperature, humidity, and cloud fraction differences 
shown Figure~\ref{fig:macmic_vs_other_presheight} reveal that:
\begin{itemize}
\item The warming and decreases in cloud fraction {\it around 850~hPa in the subtropics} 
(Figure~\ref{fig:macmic_vs_other_presheight}a and j)
are primarily attributable to shorter step sizes {\it outside} 
the cloud macro- and microphysics subcycles (Figure~\ref{fig:macmic_vs_other_presheight}c and l);
\item The cooling, drying, and cloud fraction decreases in the {\it tropical middle and upper troposphere}
(Figure~\ref{fig:macmic_vs_other_presheight}a, g, and j) 
are attributable to shortened $\Delta t_{\rm macmic}$
(Figure~\ref{fig:macmic_vs_other_presheight}b, h, and k) ;
\item 
The decreases in cloud fraction in the 
{\it mid-latitude near-surface layers} 
are also attributable to shortened $\Delta t_{\rm macmic}$
(Figure~\ref{fig:macmic_vs_other_presheight}j and k).
\end{itemize}
%
%

%---- FIGURE ---
%--------------

%---- FIGURE ---
%----------------

%---- FIGURE ---
%----------------

Geographical distributions of high-cloud and low-cloud fraction changes 
are shown in Figure~\ref{fig:macmic_vs_other_maps_f}.
The corresponding LW, SW, and total CRE changes are shown in 
Figure~\ref{fig:macmic_vs_other_maps_CRE}.
Consistent with the signatures seen in the pressure-latitude cross-sections 
in Figure~\ref{fig:macmic_vs_other_presheight},
one can see the major impact of $\Delta t_{\rm macmic}$ on 
high-cloud fraction (Figure~\ref{fig:macmic_vs_other_maps_f} top row)
and LWCRE (Figure~\ref{fig:macmic_vs_other_maps_CRE} top row).
The step sizes outside the
cloud macro- and microphysics subcycles play a major role 
in affecting the low-cloud fraction (Figure~\ref{fig:macmic_vs_other_maps_f} second row)
and SWCRE (Figure~\ref{fig:macmic_vs_other_maps_CRE} second row).
Although reductions in the various step sizes 
all lead to weakening of both LWCRE and SWCRE,
the total CRE changes seen in Figure~\ref{fig:macmic_vs_other_maps_CRE}g 
are dominated by the SW changes 
attributable to reduced low-cloud fractions
associated with shorter time steps outside the
cloud macro- and microphysics subcycles 
(Figures~\ref{fig:macmic_vs_other_maps_CRE}f and Figure~\ref{fig:macmic_vs_other_maps_f}f).

These results might appear counter-intuitive at first glance.
Since the tropical upper troposphere is strongly affected by deep convection
and the resulted detrainment of water vapor and cloud condensate,
one might have assumed the sensitivities in these 
regions to be caused primarily by step sizes associated 
with the deep convection parameterization -- or dynamics and other processes that
introduce atmospheric instability which in turn triggers deep convection.
Yet the results shown in Figures~\ref{fig:macmic_vs_other_presheight} and \ref{fig:macmic_vs_other_maps_f}
suggest that the cloud fraction decreases in these regions
are caused by shortening $\Delta t_{\rm macmic}$, 
the step size used by turbulence, shallow convection, and stratiform clouds.
A separate study has found evidence that 
the sensitivities in the tropical upper troposphere 
have to do with the representation of ice cloud microphysics in EAM. 
Prior work, e.g. \citet{Hardiman_2015_MetUM_TTL_TQ},
showed that the sedimentation and depositional growth of ice particles
can directly affect humidity in this region, while the optical properties and abundance of
ice crystals can affect SW and LW radiation 
and hence temperature in the upper troposphere; how $\Delta t_{\rm macmic}$
affects those physical processes in EAM will be investigated in follow-up work.
The link between mid-latitude near-surface clouds and 
$\Delta t_{\rm macmic}$ is unclear and needs further exploration.

In the tropical and subtropical lower troposphere, 
$\Delta t_{\rm macmic}$ appears to 
have, as expected, significant impacts on 
humidity (Figure~\ref{fig:macmic_vs_other_presheight}e and h),
cloud fraction (Figure~\ref{fig:macmic_vs_other_maps_f}e), and CRE
(Figure~\ref{fig:macmic_vs_other_maps_CRE}e and h).
On the other hand, for low-cloud fraction and CRE, 
the sensitivities to the step sizes used {\it outside} the 
cloud macro- and microphysics subcycles 
turn out to be substantially stronger
(Figure~\ref{fig:macmic_vs_other_maps_f}f, \ref{fig:macmic_vs_other_maps_CRE}f, and \ref{fig:macmic_vs_other_maps_CRE}i).
This suggests that
the low-cloud differences between v1\_All\_Shorter and v1\_CTRL are primarily  
manifestations of the responses of the subcycled processes 
to changes in the atmospheric environment passed into the subcyles.
In other words, hypothesis 2 is valid for the subtropical low clouds.
Next, we demonstrate in Section~\ref{sec:dynamic_vs_thermo}
that those low-cloud changes are associated with 
changes in the thermodynamic (instead of dynamic) 
features of the atmospheric environment. 
In Section~\ref{sec:further_attribution}, 
additional sensitivity experiments are presented to further attribute 
these changes to specific processes and step sizes
in the rest of EAMv1.

%====================
%\clearpage

\subsection{Dynamic versus thermodynamic responses of the subtropical climate}
\label{sec:dynamic_vs_thermo}

%---- FIGURE ---
\begin{figure}[htbp]
\centering\vspace{9mm}
\includegraphics[width=0.4\textwidth]{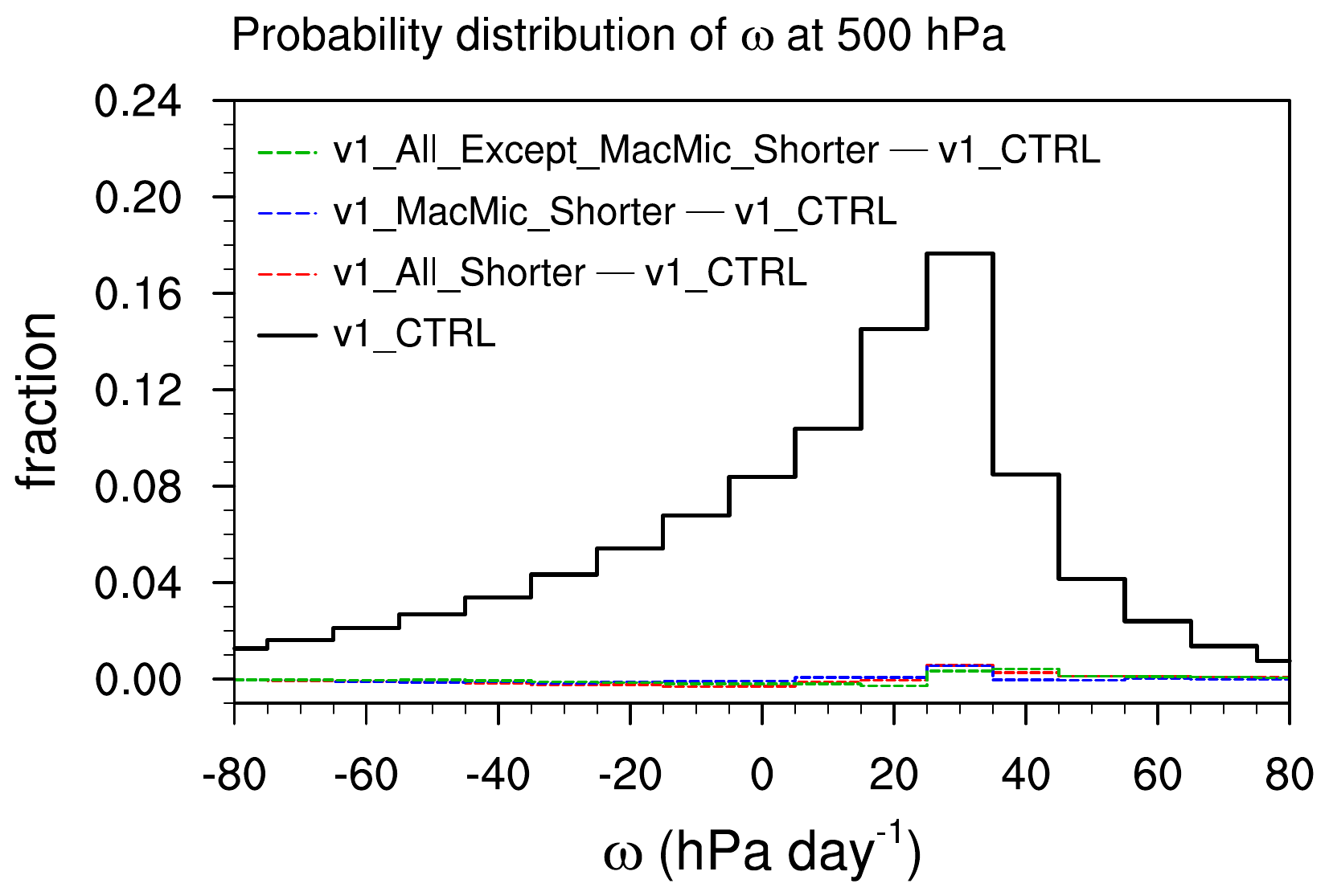}
\caption{
Frequency of occurrence of circulation regimes defined by monthly-mean 
500~hPa vertical velocity $(\omega)$ in the latitude band of 35$^\circ$S--35$^\circ$N. 
Solid black line shows the probability distribution in v1\_CTRL. 
Dashed color lines show differences in the probability distribution between other simulations and v1\_CTRL.
}\vspace{-2mm}
\label{fig:EAMv1_dynrgm_omega_2d}
\end{figure} 
%--------------

%---- FIGURE ---
\begin{figure*}[htbp]
\centering
\includegraphics[width=0.95\textwidth]{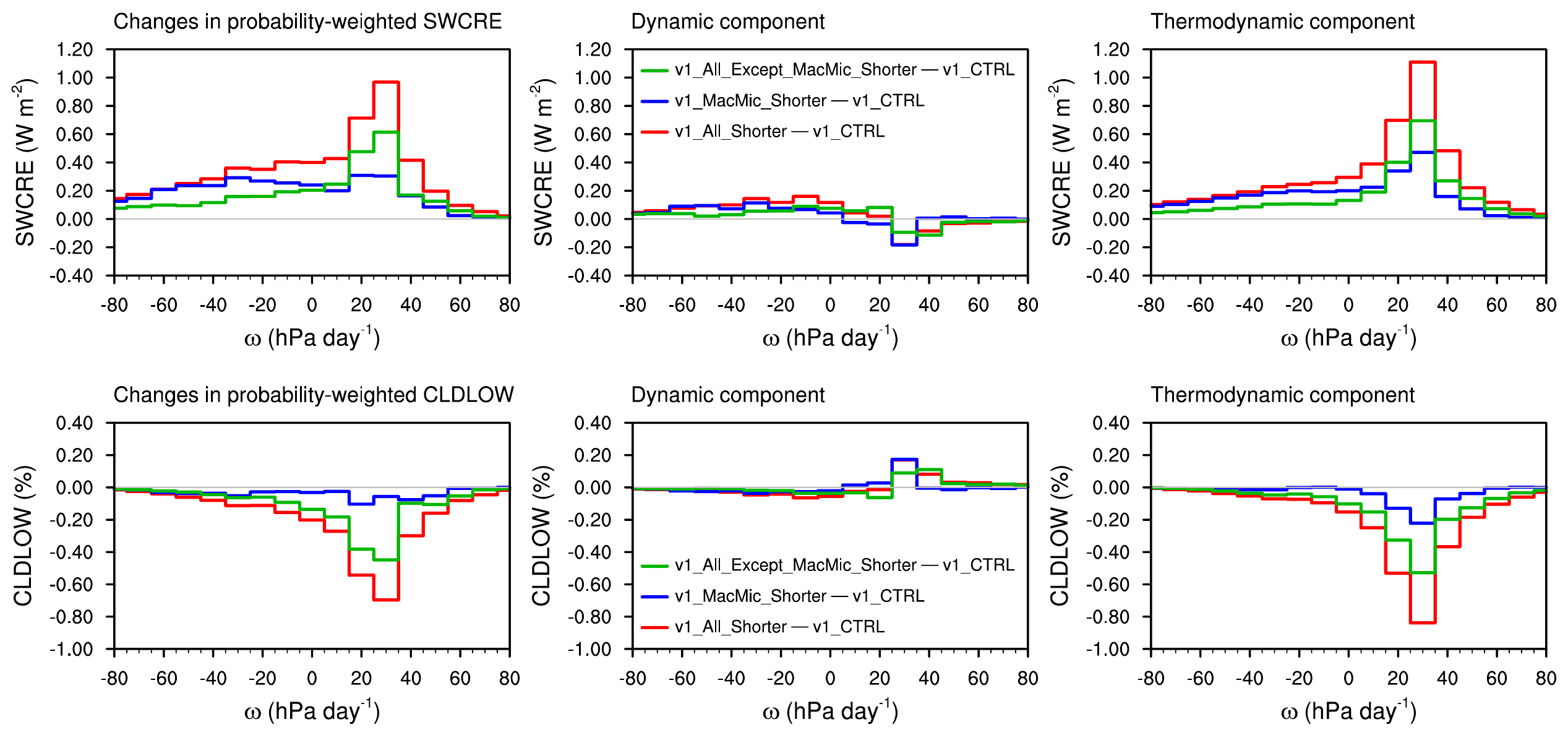}
\caption{
Left column: changes in the probability-weighted SWCRE (upper row) and low-cloud fraction (lower row) in circulation regimes characterized by grid-resolved 500~hPa $\omega$ (cf. definitions in Eqs.~\ref{eq:probability_weighted_psi} and \ref{eq:decomposition}). Middle and right columns: the dynamic and thermodynamic components of the changes (cf. Eq.~\ref{eq:decomposition}). 
Details of the analysis can be found in Section~\ref{sec:dynamic_vs_thermo}.
}
\label{fig:EAMv1_dynrgm_selvar}
\end{figure*} 
%--------------

Large-scale subsidence is one of the key features of the subtropical climate. To find out whether the reduced low-cloud fraction and weaker CRE in v1\_All\_Shorter are associated with weakened subsidence, the method from \citet{Bony_et_al:2004} is used to compare the dynamic and thermodynamic components of the low-cloud changes.

We first examined the geographical distribution of grid-resolved vertical velocity $\omega$ at 500~hPa; 
the differences among the various simulations discussed so far appeared to be rather small 
and statistically insignificant, and hence are not shown here. 
The conclusion of insignificant changes in vertical velocity can also be 
inferred from the frequency of occurrence of 500~hPa $\omega$, 
denoted here as $P_\omega$, shown in Figure~\ref{fig:EAMv1_dynrgm_omega_2d}. 
Here, $P_\omega$ is diagnosed using monthly-mean grid-point-by-grid-point $\omega$ values 
in the latitude band of 35$^\circ$S to 35$^\circ$N. 
The solid black line in Figure~\ref{fig:EAMv1_dynrgm_omega_2d} is $P_\omega$ in v1\_CTRL;
the dashed color lines show differences in $P_\omega$ between other simulations and v1\_CTRL. 
The differences appear to be close to zero compared to $P_\omega$ in v1\_CTRL.

We then followed \citet{Bony_et_al:2004} and defined circulation regimes using monthly mean $\omega$. For a circulation regime associated with $\omega$ values between $\omega_1$ and $\omega_2$, we refer to the integral of a generic physical quantity $\psi$ weighted by the probability density function $p(\omega)$ as the probability-weighted $\psi$, i.e.,
\begin{equation}
{\psi_{(\omega_1,\omega_2)}} = \int_{\omega_1}^{\omega_2} \psi\, p(\omega)\, d\omega\,.
\label{eq:probability_weighted_psi}
\end{equation}
Following \citet{Bony_et_al:2004}, changes in the probability-weighted $\psi$ can be decomposed as follows:
\begin{eqnarray}
\Delta{{\psi_{(\omega_1,\omega_2)}}} 
&\approx&
    \underbrace{\int_{\omega_1}^{\omega_2} \psi\, \Delta\!\left[p(\omega)\right]\, d\omega}_{\rm dynamic} 
  + \underbrace{\int_{\omega_1}^{\omega_2} \left(\Delta\psi\right) p(\omega) d\omega}_{\rm thermodynamic} \nonumber\\
&& + \underbrace{\int_{\omega_1}^{\omega_2} \left(\Delta\psi\right) \Delta\!\left[p(\omega)\right] d\omega}_{\rm covariation}\,
\label{eq:decomposition}
\end{eqnarray}
In Figure~\ref{fig:EAMv1_dynrgm_selvar}, we present changes in the 
probability-weighted low-cloud fraction and SWCRE in the left column, 
and their decomposition in the middle and right columns.
The results suggest that the low-cloud fraction and SWCRE changes in regions associated with subsidence 
can be attributed primarily to the thermodynamic responses of 
the model atmosphere instead of vertical velocity changes.

%============================
\subsection{Further attribution of subtropical low-cloud changes}
\label{sec:further_attribution}

In earlier sections, it has been shown that the reduction in subtropical marine 
low-cloud fraction and CRE in v1\_All\_Shorter are caused primarily by the 
use of shorter time steps 
for model components and their coupling (i.e., information exchange) 
outside the cloud macro- and microphysics subcycles.
We now make an attempt to refine the granularity of the attribution.
Additional sensitivity experiments are discussed in this subsection and 
summarized as group III in Tables~\ref{tab:exps} and \ref{tab:exps_namelist}.
An overview of the attribution process is provided in Figure~\ref{fig:attribution}
with pointers to the figures that show the results.

%---- FIGURE ---
\begin{figure}[htbp]
\centering
\includegraphics[width=0.4\textwidth]{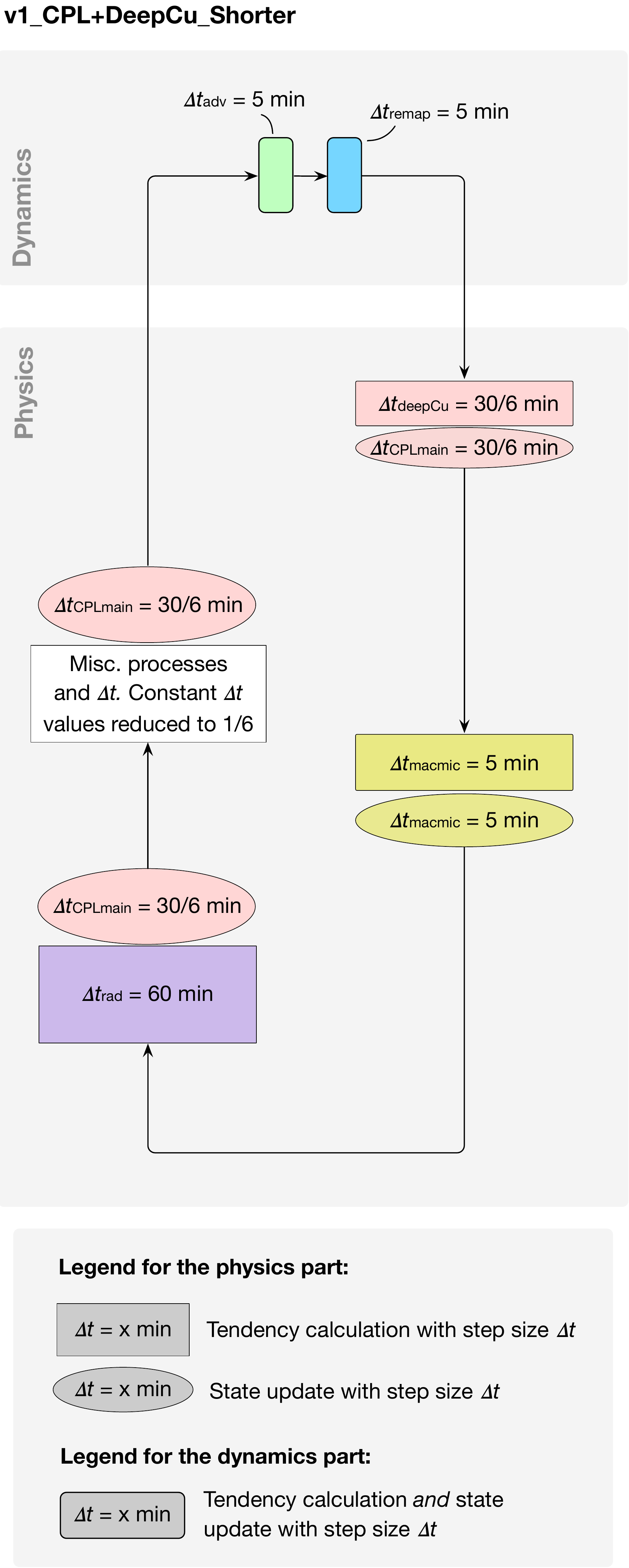}
\caption{
Schematic showing time step sizes used in simulation v1\_CPL+DeepCu\_Shorter. 
The color coding follows Figure~\ref{fig:ordering}a.
Further details can be found in Section~\ref{sec:dyn_rad}.
The simulation setup is summarized in 
Tables~\ref{tab:exps} and \ref{tab:exps_namelist}.
}
\label{fig:schematic_CPL+DeepCu}
\end{figure} 

%---- FIGURE ---
\begin{figure}[htbp]
\centering\vspace{1cm}
\includegraphics[width=0.38\textwidth]{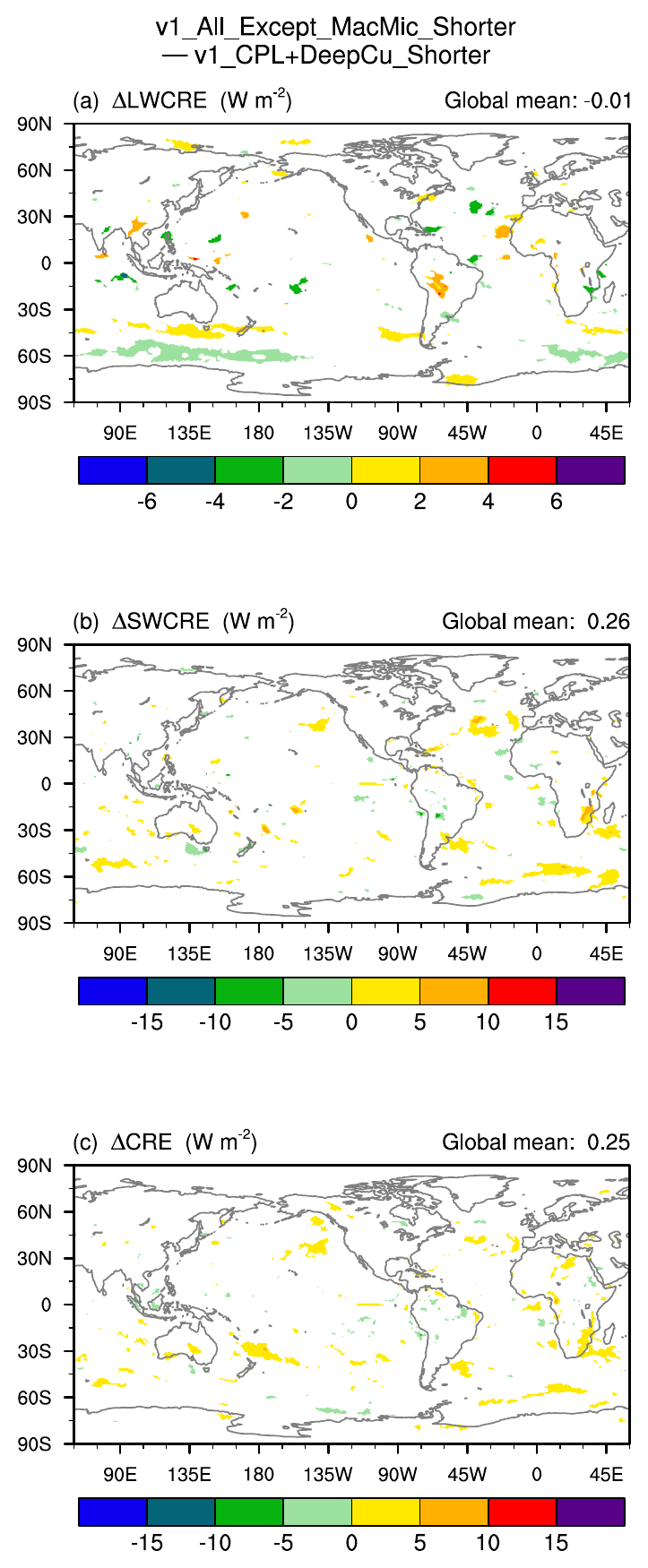}
 \caption{
10-year mean CRE differences between simulations
v1\_All\_Except\_MacMic\_Shorter (cf. schematic in Figure~\ref{fig:schematic_macmic_vs_other}b) and 
v1\_CPL+DeepCu\_Shorter (cf. schematic in Figure~\ref{fig:schematic_CPL+DeepCu}),
revealing the impact of shortened dynamics and radiation time steps. 
White indicates statistically insignificant differences.
The simulation setups are summarized in Tables~\ref{tab:exps} and \ref{tab:exps_namelist}.
}
\vspace{-2mm}
\label{fig:rad_dyn_CRE_maps}
\end{figure}
%--------------

\subsubsection{Resolved dynamics and radiation} 
\label{sec:dyn_rad}

A simulation labeled as {\bf v1\_CPL+DeepCu\_Shorter} 
in Tables~\ref{tab:exps} and \ref{tab:exps_namelist}
(cf. schematic in Figure~\ref{fig:schematic_CPL+DeepCu})
was configured to be the same as  
v1\_All\_Except\_MacMic\_Shorter (schematic in Figure~\ref{fig:schematic_macmic_vs_other}b)  
except that $\Delta t_{\rm adv}$ 
(the horizontal advection time step in the dynamical core)
and $\Delta t_{\rm rad}$ (the interval for calculating radiative cooling/heating rates) 
are reverted to their default values in v1\_CTRL\footnote{
Because of the required relationship among $\Delta t_{\rm remap}$, 
$\Delta t_{\rm adv}$ and $\Delta t_{\rm CPLmain}$ (cf. Section~\ref{sec:model}), 
this new simulation ended up using $\Delta t_{\rm remap}$=~5~min
which fell between what was used in v1\_All\_Except\_MacMic\_Shorter (15/6~=~2.5~min)
and v1\_CTRL (15~min), but the effect is expected to be small.}.
The comparison of this pair of simulations 
reveals the impact of $\Delta t_{\rm adv}$  and $\Delta t_{\rm rad}$
shown in Figure~\ref{fig:rad_dyn_CRE_maps}.
The CRE differences appear to have small and mostly insignificant magnitudes.
The small yet systematic differences in LWCRE in the Southern Hemisphere 
midlatitudes (Figure~\ref{fig:rad_dyn_CRE_maps}a) 
indicate a shift in the location of the storm tracks, 
but no systematic signals are seen in the lower latitudes.
Therefore we conclude that the impact of dynamics and radiation time steps on subtropical clouds is small, 
at least in the context of the currently used process ordering and splitting/coupling methods.

\subsubsection{Coupling between cloud macro/microphysics and other processes}
\label{sec:cpl}

%---- FIGURE ---
\begin{figure}[htbp]
\centering
\includegraphics[width=0.4\textwidth]{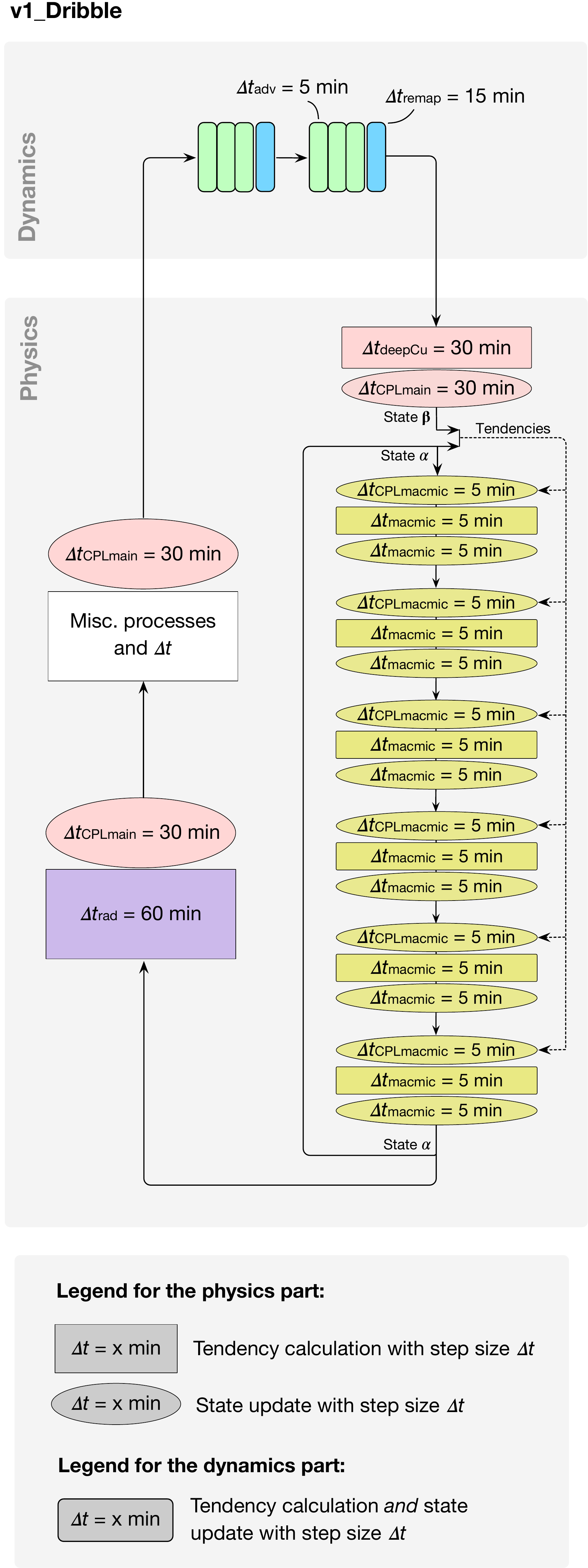}
\caption{
Schematic showing time step sizes and the sequence of calculations 
used in simulation v1\_Dribble.
The color coding follows Figure~\ref{fig:ordering}a.
The simulation setup is summarized in  
Tables~\ref{tab:exps} and \ref{tab:exps_namelist}.
The results are discussed in Section~\ref{sec:cpl}.
}
\label{fig:schematic_dribble}
\end{figure}
%--------------

%---- FIGURE ---
\begin{figure*}[htbp]
\centering
\includegraphics[width=0.7\textwidth]{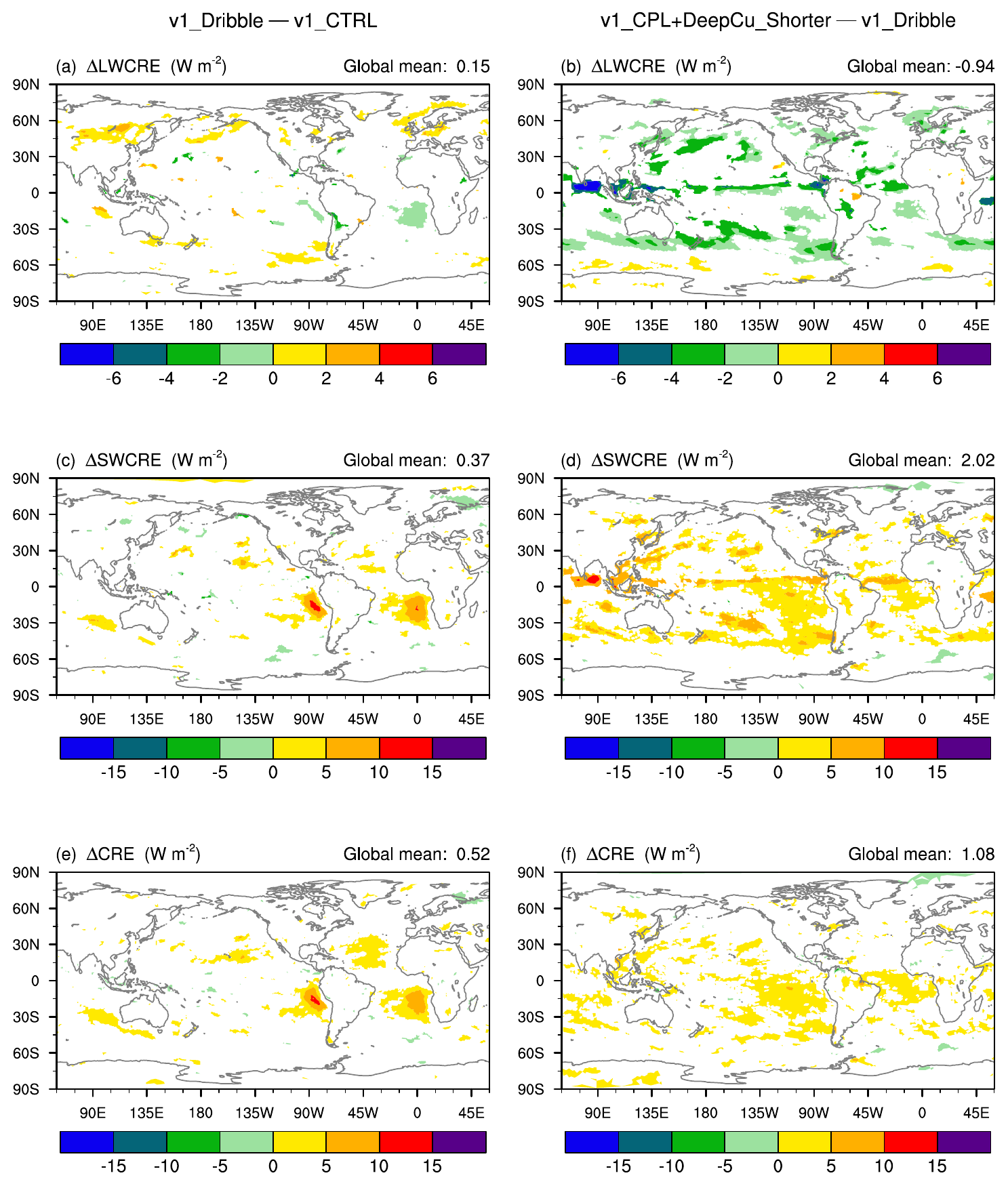}
 \caption{
Left: differences between 
v1\_Dribble (cf. schematic in Figure~\ref{fig:schematic_dribble}) and
v1\_CTRL (cf. schematic in Figure~\ref{fig:schematic_v1_CTRL_and_All_Shorter}),
revealing the impact of coupling between the subcycled cloud macro-/microphysics and the rest of EAM.
Right: differences between 
v1\_CPL+DeepCu\_Shorter (cf. schematic in Figure~\ref{fig:schematic_CPL+DeepCu}) and 
v1\_Dribble (cf. schematic in Figure~\ref{fig:schematic_dribble}),
revealing the impact of step sizes used by various other parameterizations 
(deep convection, gravity wave drag, miscellaneous aerosol processes) and the coupling among them.
White indicates statistically insignificant differences.
The simulation setups are summarized in Tables~\ref{tab:exps} and \ref{tab:exps_namelist}.
 }
\vspace{-2mm}
\label{fig:dribble_vs_other}
\end{figure*}
%-------------

Since Section~\ref{sec:dyn_rad} has shown that 
the step sizes of resolved dynamics and radiation time steps have only very limited impacts, 
we are left with two step sizes to explore, $\Delta t_{\rm CPLmain}$ and $\Delta t_{\rm deepCu}$,
to answer the question which step sizes outside the cloud macro- and microphysics subcycles
are responsible for the subtropical CRE changes shown in the rightmost column of Figure~\ref{fig:macmic_vs_other_maps_CRE}.
As explained in Section~\ref{sec:model} and illustrated by color coding 
in Figure~\ref{fig:schematic_v1_CTRL_and_All_Shorter}a, these two step sizes have the same value in EAMv1, 
and the single $\Delta t_{\rm CPLmain}$ also controls the coupling frequency among the majority of 
the parameterizations as well as between physics and dynamics. 
This makes further attribution somewhat difficult unless changes are made to the model source code. 
Nevertheless, our exploration revealed that the coupling between 
the subcycled cloud parameterizations and the rest of the model was impactful. 

Figure~\ref{fig:schematic_dribble} shows the schematic for 
a simulation called {\bf v1\_Dribble} which uses EAMv1's default step sizes for 
all the individual model components but  
a revised scheme for the coupling between the stratiform cloud subcycles and the rest of the model. 
In the revised scheme, the atmospheric temperature, specific humidity, 
as well as cloud liquid and ice concentrations that are passed to the first 
cloud macro/microphysics subcycle are {\it no longer} the values updated by deep convection and dynamics etc.
(i.e., {\it no longer} ``state $\beta$'' in Figure~\ref{fig:schematic_dribble}).
Instead, the older snapshot saved after the last (i.e., 6th) 5~min cloud macro/microphysics subcycle 
in the previous main time step
(``state $\alpha$'' in Figure~\ref{fig:schematic_dribble})
is provided together with the tendencies 
caused by all processes between points $\alpha$ and $\beta$ in the schematic. 
At the beginning of each subcycle, those tendencies are used to update the 
atmospheric state using a step size of 5~min, 
as illustrated by the greenish-yellow ovals labeled with ``$\Delta t_{\rm CPLmacmic}$~=~5~min'' in Figure~\ref{fig:schematic_dribble}. 
This ``dribbling'' method is conceptually similar to the physics-dynamical coupling scheme
used by EAMv1's dynamical core for temperature and horizontal winds
\citep{Zhang_et_al:2018,Rasch_et_al:2019}, 
and also similar to the two other instances of tendency-involved
process coupling mentioned at the end of Section~\ref{sec:EAMv1}.
This ``dribbling'' can be viewed as an example of the sequential-tendency splitting method 
defined in \citet{Donahue_Caldwell:2018}.
To help distinguish this ``dribbling" from the original splitting method 
depicted in Figure~\ref{fig:schematic_v1_CTRL_and_All_Shorter}a,
we introduced the notation $\Delta t_{\rm CPLmacmic}$ in Eq.~\eqref{eq:dt_macmic} 
and Figure~\ref{fig:schematic_dribble} as well as in
Table~\ref{tab:exps} and Figure~\ref{fig:attribution}.
$\Delta t_{\rm CPLmacmic}$~=~30~min in v1\_CTRL and
5~min in v1\_Dribble.

``Dribbling" provides a more frequent coupling from the perspective of
the subcycled cloud macro/microphysics, while the feedback to the 
processes outside the subcycles still occurs at longer intervals of $\Delta t_{\rm CPLmain}$.
A detailed explanation of the motivation for this ``dribbling" and 
an in-depth analysis of its impact on the atmospheric water budget 
will be the topic of a separate paper. Here we only show the CRE differences 
between the two simulations v1\_Dribble and v1\_CTRL in the left column of Figure~\ref{fig:dribble_vs_other}.
Weakened SWCRE and total CRE are found
over the eastern parts of the subtropical oceans, 
especially in the Peruvian and Namibian stratocumulus regions.
This suggests that the strongest local reduction of SWCRE and/or total CRE 
seen earlier in simulation v1\_All\_Except\_MacMic\_Shorter (Figure~\ref{fig:macmic_vs_other_maps_CRE}f and i)
are primarily attributable to more frequent coupling between the subcycled cloud macro/microphysics
and the rest of EAM.

%====================
\subsubsection{Deep convection}
\label{sec:deep_cu}

We now attempt to attribute the time step sensitivities seen in simulation 
v1\_All\_Except\_MacMic\_Shorter (Figure~\ref{fig:macmic_vs_other_maps_CRE}, right column)
that are not explained by ``dribbling'' (Figure~\ref{fig:dribble_vs_other}, left column)
or the time steps of dynamics and radiation (Figure~\ref{fig:rad_dyn_CRE_maps}).
This can be done by comparing the simulation v1\_CPL+DeepCu\_Shorter introduced in Section~\ref{sec:dyn_rad} 
and v1\_Dribble discussed in Section~\ref{sec:cpl}, as the two experiments 
share the same $\Delta t_{\rm adv}$, $\Delta t_{\rm rad}$, and $\Delta t_{\rm CPLmacmic}$
while they differ in the step sizes used for deep convection and it's coupling to 
other processes, as well as miscellaneous processes like 
land surface, gravity wave drag, aerosols, and the coupling among them.

The right column in Figure~\ref{fig:dribble_vs_other} shows the 10-year mean differences in CRE,
revealing weaker LWCRE and SWCRE along the equatorial ITCZ 
(Inter Tropical Convergence Zone, where deep convection is important) 
and in the subtropics and equatorward flanks of the storm tracks.
The LW and SW changes largely cancel each other along the equator 
and near the storm tracks, leaving differences in the net CRE visible only in 
the trade cumulus regions. 
The signatures of a net cancellation in LWCRE and SWCRE along the ITCZ provide hints that 
there is a change in behavior in the deep convection regime.

Similar to the discussions in earlier sections on the stratiform cloud parameterizations,
time step sensitivities associated with the deep convection parameterization can 
potentially be caused by the temporal truncation errors inside the parameterization
or the coupling between the parameterization and other model components, or both.
In the default EAMv1 and its recent predecessors, no subcycles are used for deep convection, 
meaning that the convection 
time step $\Delta t_{\rm deepCu}$ and coupling time step $\Delta t_{\rm CPLmain}$
are tied together. Therefore, without further code modifications and simulations, 
we can not yet further attribute the sensitivities seen in the right column of 
Figure~\ref{fig:dribble_vs_other}.
\citet{Williamson:2013} discussed how the interplay between 
convection and stratiform clouds can be affected by their corresponding
timescales and the model time step. Based on that study,
one can speculate that process coupling might be an important 
cause of the sensitivities seen in the right column of 
Figure~\ref{fig:dribble_vs_other}. 
Whether this is indeed the case needs
to be verified in future studies. 
Here we only make a brief comment that while the 
results in \citet{Williamson:2013} are commonly interpreted as
the deep convection parameterization being constrained by the assumed timescale,
our preliminary exploration described in Appendix~\ref{sec:dt_tau}
suggests that the timestep-timescale ratio alone cannot 
explain the changes in CRE shown in 
the right column of Figure~\ref{fig:dribble_vs_other}.
There are other significant factors related to process coupling 
that need to be identified and understood in the future.

\conclusions  %% \conclusions[modified heading if necessary]
\label{sec:conclusions}

%---- FIGURE ---
\begin{figure*}[htbp]
\centering
\includegraphics[width=1\textwidth]{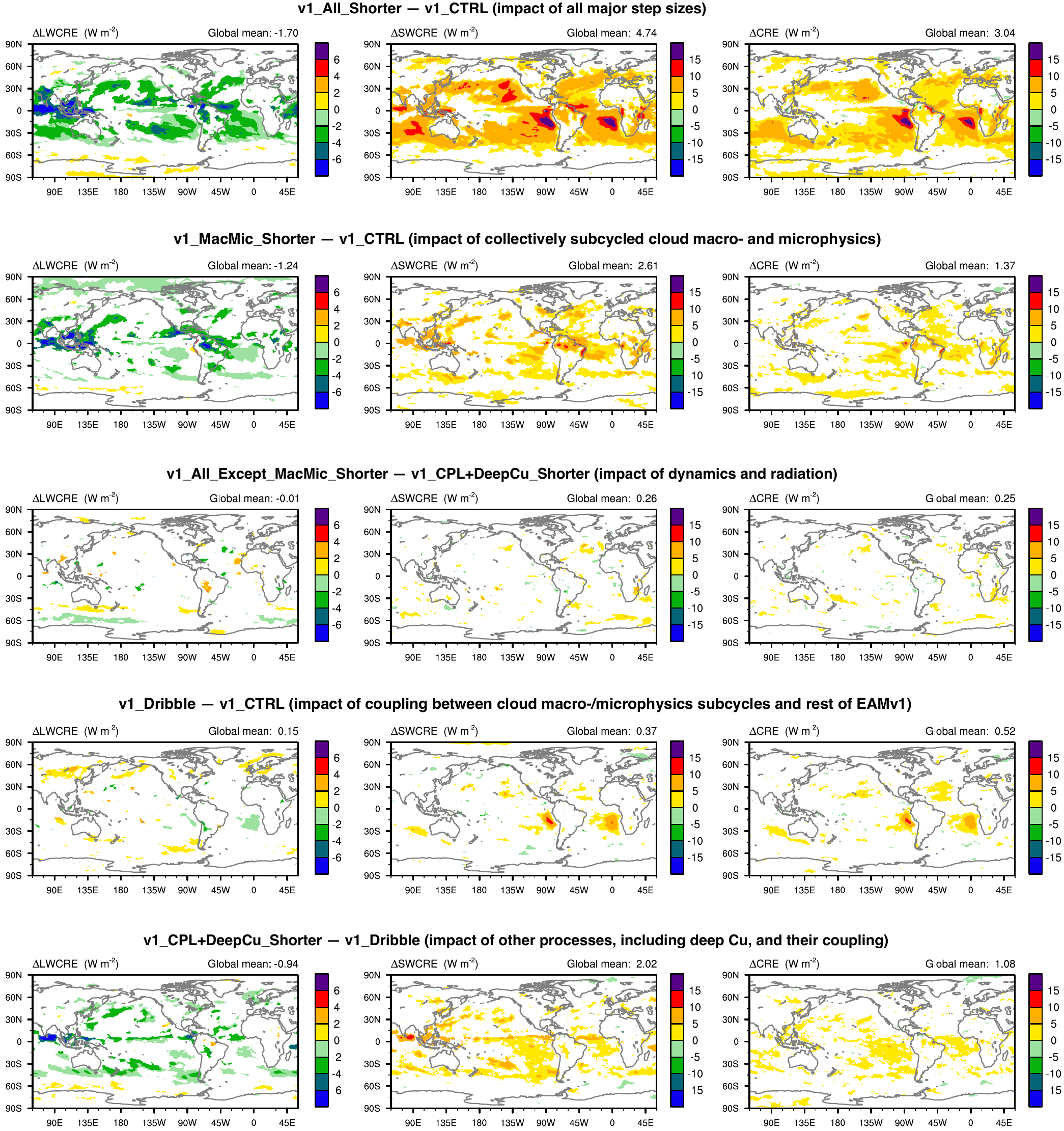}
 \caption{
Attribution of the 10-year mean CRE differences between v1\_All\_Shorter and v1\_CTRL (first row)
to various components of EAMv1 (lower rows).
Left column: LWCRE; middle column: SWCRE; right column: total CRE.
White indicates statistically insignificant differences.
The attribution process is summarized in Figure~\ref{fig:attribution}.
The simulation setups are summarized in Tables~\ref{tab:exps} and \ref{tab:exps_namelist}.
Schematics depicting EAM's time integration loop and 
different step sizes can be found in 
Figures~\ref{fig:schematic_v1_CTRL_and_All_Shorter}, 
\ref{fig:schematic_macmic_vs_other}, 
\ref{fig:schematic_CPL+DeepCu}, and
\ref{fig:schematic_dribble}.
 }
\vspace{-2mm}
\label{fig:attribution_all_together}
\end{figure*}

This study evaluated the strength of time step sensitivities in 10-year present-day climate simulations conducted with the EAMv1 atmospheric model at 1-degree horizontal resolution. A proportional, factor-of-6 reduction of time step size in major components of the model (simulations v1\_All\_Shorter versus v1\_CTRL) 
was found to result in changes in the long-term mean climate that were significant both statistically and physically. A systematic warming was found in the low-latitude areas in the near-surface levels and a systematic cooling was seen aloft, with 10-year zonal mean temperature differences of up to 0.5~K. 
The zonal mean relative humidity was found to decrease by 1\%--10\% throughout the troposphere. 
Sizable zonal mean cloud fraction decreases were seen 
at most latitudes in the upper troposphere (10\%--20\%), 
in the subtropical lower troposphere (more than 20\%), and
in the mid-latitude near-surface layers (10\%--20\%).
In terms of geographical distribution, the most pronounced annual mean changes are the decreases in 
total cloud cover (10\%--50\%) 
and CRE (20\%--50\%) over the subtropical marine stratocumulus and trade cumulus regions.
The global mean CRE weakens by about 3~W~m$^{-2}$, corresponding to a relative decrease of 12\%.

The comparison of model results with a comprehensive set of observational data indicated 
that the changes caused by step size reduction led to a degradation in model fidelity, 
in terms of both the global mean statistics and the geographical distributions. 
Although this is not surprising given the careful tuning EAMv1 has undergone, 
the compensation of time integration error by parameter tuning or by other sources 
of model error is undesirable.
This compensation implies the need for additional tuning 
to achieve a new compensation when model time steps are shortened for high-resolution simulations. 
It would be more desirable to identify time-stepping algorithms with numerical errors 
that are small enough that the simulation fidelity is insensitive to 
reasonable variations in step size; that is, so that the simulation quality 
is determined by physical understanding (or lack of it).

In order to provide clues for future efforts on reducing time-stepping errors in EAM, 
additional simulations were conducted to tease out some of the sources of time step sensitivities seen in EAMv1. 
Most of those simulations made use of flexible choices of time step sizes currently available 
in various subsets of EAM's components. 
One of the simulations (v1\_Dribble) used an alternate numerical scheme to couple the collectively 
subcycled shallow cumulus and stratiform cloud macro/microphysics parameterizations with the rest of EAM
at a higher frequency.
A simulation discussed in Appendix~\ref{sec:dt_tau} 
used a different value for the CAPE removal timescale in the 
deep convection parameterization to investigate the impact of the ratio of time step to this timescale.

Analysis of the results focused primarily on the annual mean cloud fraction and CRE.
We found that the most notable sensitivity in the simulations was 
changes in total cloud cover and CRE in the subtropical marine stratocumulus 
and trade cumulus regimes.
Our analysis revealed that this sensitivity was {\it not} caused primarily by the step size 
used for treating some of the most important processes in those regimes 
(turbulence, shallow cumulus and stratiform cloud macro- and microphysical processes, see simulation v1\_MacMic\_Shorter),
but rather by the strategy used to couple those processes to other components of the model
(see simulation v1\_Dribble).  
On the other hand, the step size of the cloud macro- and microphysics subcycles
had quite an important impact on cloud fraction at most latitudes 
in the upper troposphere between 100~hPa and 400~hPa,
as well as in the mid-latitude near-surface layers.
Additional simulations and analysis 
revealed that the deep convection parameterization and its coupling with other processes
significantly affected trade cumulus.
Impacts of the step sizes used by the dynamical core and radiation were small.
In Figure~\ref{fig:attribution_all_together}, we have reorganized some of the 
CRE difference plots presented in earlier sections: a different panel layout is used
to facilitate a direct comparison of the impacts of step sizes used by different model components.
Recent follow-up or independent studies have provided 
insights into the impact of process coupling on marine stratocumulus clouds
and the impact of macro/microphysics time step on ice cloud formation.
Those results will be reported in separate papers.
The mechanisms behind the other sensitivities shown in the figure 
still need to be investigated.

Using the analysis method of \citet{Bony_et_al:2004}, 
we found that the subtropical low-cloud changes were primarily local, 
thermodynamic responses of the model atmosphere while the impact of 
circulation (vertical velocity) changes was very small.
This conclusion has practical implications for follow-up investigations: 
since circulation changes are negligible and local cloud processes are fast, 
it should be feasible to use nudged 1-year simulations 
\citep{Kooperman:2012,Zhang_et_al:2014,Sun:2019} 
or even ensembles of few-day simulations \citep{Xie:2012,ma:2013a,Wan_et_al:2014} 
to help carry out further investigations at process level
and meanwhile keep the numerical experiments computationally economical.

Coincidentally, when our manuscript was submitted to {\it Geoscientific Model Development},
a paper by \citet{Santos_2020_time_step_sensitivity} was submitted to 
a different journal which described an independent study that also 
attempted to quantify and attribute
time step sensitivities in EAMv1. 
Their experimental strategy and ours turned out to be similar, 
although the details differed.
Their analysis had a stronger focus on global mean precipitation rates 
and zonal mean cloud amounts whereas
our attribution  focused more on the geographical 
distribution of CRE.

While both this study and the work of \citet{Santos_2020_time_step_sensitivity}
focused on one specific AGCM, it would be useful to carry out
similar exercises with other models.
Because of considerations of computational cost,
numerical models used for operational weather forecasts 
and climate research generally tend to use the longest step sizes
that would provide satisfactory results. The chosen step sizes,
however, influence key simulated features to an extent that
is not always clear. If a time step sensitivity quantification
exercise like ours presented in Section~\ref{sec:quantification}
reveals strong sensitivities, that would provide a motivation 
to understand the causes of the sensitivities, and, in the next step,
revise the numerical methods to provide higher accuracy without 
substantially increasing the computational cost.

%% The following commands are for the statements about the availability of data sets and/or software code corresponding to the manuscript.
%% It is strongly recommended to make use of these sections in case data sets and/or software code have been part of your research the article is based on.

\codeavailability{
The EAMv0 and v1 source codes and run scripts used in this study can be found 
on Zenodo at \url{https://doi.org/10.5281/zenodo.4118705}.
} %% use this section when having only software code available

%Model output used for analyses in the paper is temporarily stored at PNNL's Research Computational Facility under \textcolor{red}{\url{https://}}. Upon acceptance of the paper for publication, the data will be made available through the U.S. Department of Energy's Data Explorer at \url{https://www.osti.gov/dataexplorer/} with a designated DOI. A previous example of data preserved this way can be found at \url{https://www.osti.gov/dataexplorer/biblio/dataset/1561460}.

%\dataavailability{TEXT} %% use this section when having only data sets available

%\codedataavailability{} %% use this section when having data sets and software code available

%\sampleavailability{TEXT} %% use this section when having geoscientific samples available

%\videosupplement{TEXT} %% use this section when having video supplements available

\appendix

\section{Additional table and figure for Sections~\ref{sec:model}--\ref{sec:attribution}}    %% Appendix A

%\appendixfigures  %% needs to be added in front of appendix figures

%---- FIGURE ---
\begin{figure}[htbp]
\centering
\includegraphics[width=0.35\textwidth]{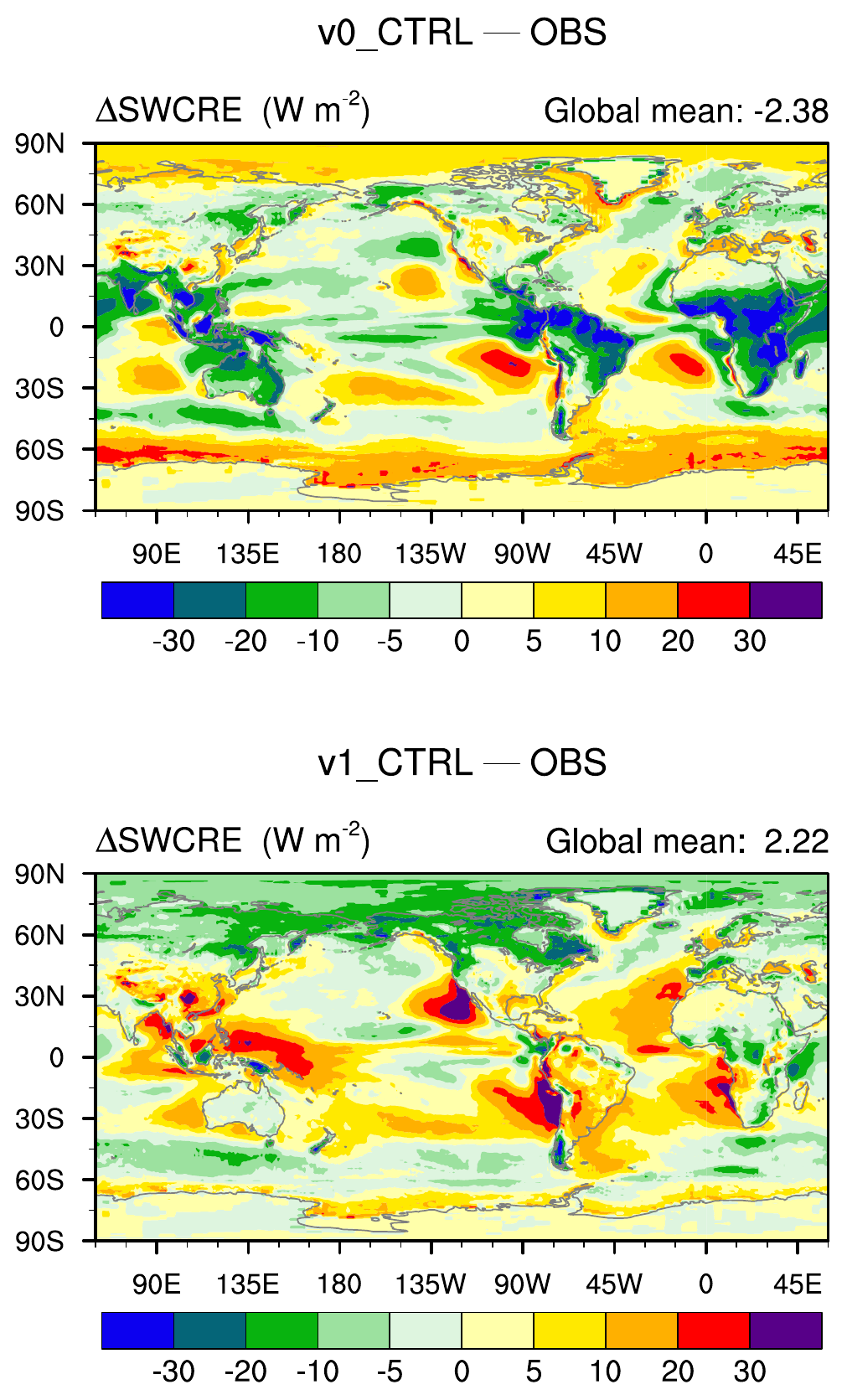}
\caption{10-year mean differences in SWCRE between simulation v0\_CTRL (top) or v1\_CTRL (bottom) 
and the 2000-2010 averages from CERES-EBAF.
}
\label{fig:SWCF_v0_v1_vs_obs}
\end{figure} 

Table~\ref{tab:exps_namelist} documents the namelist settings used in 
the EAMv0 and EAMv1 simulations presented in this paper.
Figure~\ref{fig:SWCF_v0_v1_vs_obs} presents the geographical distribution 
of SWCRE biases in v0\_CTRL and v1\_CTRL to show that the two models 
have different characteristics in the spatial distribution of model biases
(cf. Section~\ref{sec:v0}).
%

%---------------------------------------------------
%\appendixtables   %% needs to be added in front of appendix tables
%\clearpage

\begin{sidewaystable}[htbp]
\caption{Namelist setups used by the simulations listed in Table~\ref{tab:exps}. 
%The quantities given in parentheses are step sizes or timescale affected by the corresponding namelist variables.
}
\label{tab:exps_namelist}
\centering
\begin{adjustbox}{max width=0.95\textwidth}
\begin{tabular}{clll|ccccc|c}
\tophline
\multirow{2}{*}{Group} & 
\multirow{2}{*}{Simulation} &
\multirow{2}{*}{Description} &
\multirow{2}{*}{Schematic} &
\multicolumn{5}{c}{Namelist variables and their values}\\\cline{5-10}
&&&& se\_nsplit & rsplit  & dtime  
  & cld\_macmic\_num\_steps 
  & iradsw, iradlw 
  & zmconv\_tau  \\
\middlehline
0   & v0\_CTRL                        & Sect.~\ref{sec:EAMv0}           & -                               & 2   & 3     & 1800   &  N/A  & {-1, -1} & 3600 \\
\middlehline
I   & v1\_CTRL                        & Sect.~\ref{sec:EAMv1}           & Fig.~\ref{fig:schematic_v1_CTRL_and_All_Shorter}a       & 2   & 3     & 1800   &  6  & {-1, -1} & 3600\\
I   & v1\_All\_Shorter                & Sect.~\ref{sec:exps}            & Fig.~\ref{fig:schematic_v1_CTRL_and_All_Shorter}b    & 2   & 3     &  300   &  6  & {2, 2}   & 3600\\
\middlehline
II  & v1\_MacMic\_Shorter             & Sect.~\ref{sec:StCld_vs_rest}   &  Fig.~\ref{fig:schematic_macmic_vs_other}a     & 2   & 3     & 1800   & 36  & {-1, -1} & 3600\\
II  & v1\_All\_Except\_MacMic\_Shorter& Sect.~\ref{sec:StCld_vs_rest}   &  Fig.~\ref{fig:schematic_macmic_vs_other}b     & 2   & 3     &  300   &  1  & {2, 2}   & 3600 \\ 
\middlehline
III & v1\_CPL+DeepCu\_Shorter         & Sect.~\ref{sec:dyn_rad}         &  Fig.~\ref{fig:schematic_CPL+DeepCu}     & 1   & 1     &  300   &  1  & {-1, -1} & 3600\\ 
III & v1\_Dribble                     & Sect.~\ref{sec:cpl}             & Fig.~\ref{fig:schematic_dribble}& 2   & 3     & 1800   &  6  & {-1, -1} & 3600\\
\middlehline
IV & v1\_CPL+DeepCu+Tau\_Shorter     & App.~\ref{sec:dt_tau}         &  Fig.~\ref{fig:schematic_CPL+DeepCu}     & 1   & 1     &  300   &  1  & {-1, -1} & 600 \\
\bottomhline
\end{tabular}
\end{adjustbox}
\end{sidewaystable}
%--------------

\section{Deep convection timescale and time step}
\label{sec:dt_tau}

%---- FIGURE ---
\begin{figure}[htbp]
\centering
\includegraphics[width=0.35\textwidth]{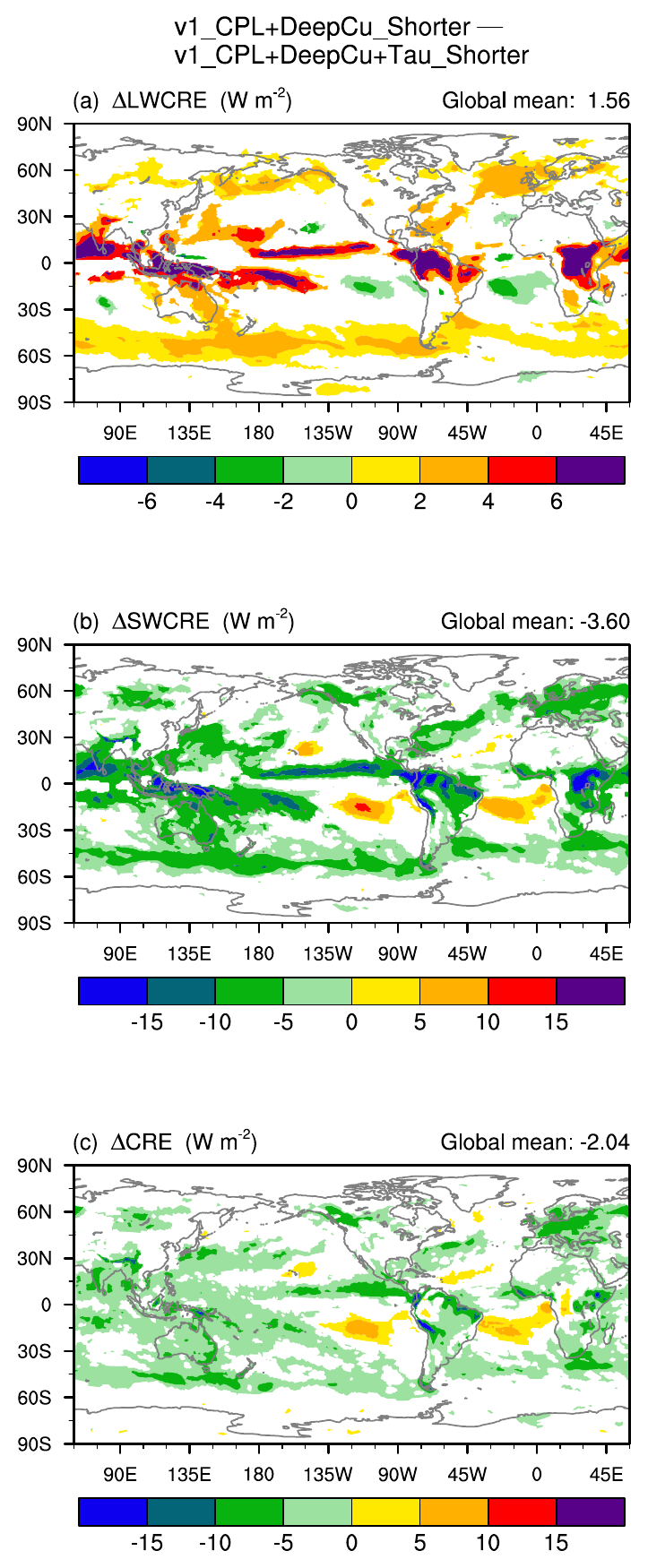}
 \caption{
10-year annual mean CRE differences between v1\_CPL+DeepCu\_Shorter and v1\_CPL+DeepCu+Tau\_Shorter,
revealing the impact of a reduced ratio of $\Delta t/\tau$ without model step size changes.
White indicates statistically insignificant differences.
The simulation setups are summarized in Tables~\ref{tab:exps} and \ref{tab:exps_namelist}.
The two simulations correspond to the same schematic shown in Figure~\ref{fig:schematic_CPL+DeepCu}.
}
\label{fig:dt_tau}
\end{figure}
%-------------

Section~\ref{sec:deep_cu} noted time step sensitivities in the deep convection regime 
but was inconclusive on the cause of such sensitivities.
Based on the work of \citet{Williamson:2013}, one can speculate that
the interactions between deep convection and stratiform cloud parameterizations
might be an important factor. In this appendix, we present 
some preliminary results to show that how those interactions
affect time step sensitivities in the deep convection regime is 
a complex topic that needs further investigation.

\citet{Williamson:2013} pointed out that convective parameterizations designed to 
remove instability and supersaturation on assumed constant $\tau$ are constrained by 
its step size in how much work such parameterizations can do in each time step. 
In contrast, large-scale condensation parameterizations designed to completely remove 
supersaturation within every time step are unconstrained, 
with implications for the column instability and depth of convection that 
in turn affects the resolved dynamical response. 
This difference in characteristic behavior can affect the simulated interactions 
between dynamics, deep convection and the stratiform cloud processes. 
\citet{Williamson:2013} showed that this time-step-time-scale issue ($\Delta t/\tau$ issue) 
could explain the occurrence of intense truncation-scale storms 
in high-resolution simulations conducted with the Community Atmosphere Model Version 4 (CAM4). 
Other studies, e.g., \citet{Mishra_et_al:2008}, \citet{Mishra_Srinivasan:2010}, 
\citet{Mishra_Sahany:2011}, \citet{Yang_2013_ZM_params}, \citet{Qian_2015_CAM5_PPE}, \citet{Yu_Pritchard:2015}, 
\citet{LinGX_2016_JAMES_nudging}, and \citet{Qian_2018_JAMES_EAMv1_short_PPE}
have also shown model sensitivities to $\Delta t$ and/or $\tau$. 

Here, it is worth noting again that in the default EAMv1 and its recent predecessors,
including CAM4 used in \citet{Williamson:2013}, 
no subcycles are used for deep convection, 
meaning that the step size used by each invocation of the convection parameterization
is the same as the step size used for the coupling of  
deep convection with other model components. 
In other words, $\Delta t$ in \citet{Williamson:2013} is both 
$\Delta t_{\rm deepCu}$ and $\Delta t_{\rm CPLmain}$ in this paper
and we have $\Delta t_{\rm deepCu}\equiv\Delta t_{\rm CPLmain}$ in these models.

It is also worth noting that the ratio of $\Delta t/\tau$ can be changed through
either the denominator or the numerator, or both.
The {\it direct} effects of varying $\tau$ are limited to 
the $\Delta t/\tau$ ratio and the strength of convective activity the ratio controls,
while a different $\Delta t$ will also change the temporal truncation error
associated with the deep convection scheme and process coupling.

The impact of a smaller $\Delta t/\tau$ ratio caused by changing only the timescale $\tau$ 
can be derived by comparing the simulation v1\_CPL+DeepCu\_Shorter introduced in Section~\ref{sec:dyn_rad} 
and a new simulation {\bf v1\_CPL+DeepCu+Tau\_Shorter}.
Both experiments used the same step sizes depicted in the schematic in Figure~\ref{fig:schematic_CPL+DeepCu}
but the values of $\tau$ differed by a factor of 6, 
hence the $\Delta t/\tau$ ratio in v1\_CPL+DeepCu\_Shorter is 1/6 of the 
ratio in v1\_CPL+DeepCu+Tau\_Shorter.
A smaller ratio decreases
the relative importance of the convective parameterization per time step
and therefore amplifies the role of the stratiform cloud parameterizations 
and associated Hadley circulation;
this makes the positive LWCRE more positive and 
the negative SWCRE more negative in convective regions.
In other words, the strengthened amplitudes of both LW and SW CRE in the ITCZ 
seen in Figure~\ref{fig:dt_tau} are
consistent with the response described in \citet{Williamson:2013}.

In contrast, the right column of 
Figure~\ref{fig:dribble_vs_other} shown in Section~\ref{sec:deep_cu}
are CRE changes corresponding to a factor-of-6 decrease in 
the $\Delta t/\tau$ ratio caused by shortening $\Delta t$;
there the signs and patterns of CRE changes in the ITCZ are different from 
what is seen in Figure~\ref{fig:dt_tau}. 
The discrepancies between the two figures
suggest that the more frequent invocation of deep convection and 
more frequent coupling with other processes
have led to consequences that compensate (in fact overcompensate) 
the impact of a smaller $\Delta t/\tau$. 
In other words, the overall responses of the annual mean CREs 
to a shortened $\Delta t$ are {\it inconsistent} with 
the timestep-timescale argument in \citet{Williamson:2013}.

\noappendix       %% use this to mark the end of the appendix section. Otherwise the figures might be numbered incorrectly (e.g. 10 instead of 1).

%% Regarding figures and tables in appendices, the following two options are possible depending on your general handling of figures and tables in the manuscript environment:

%% Option 1: If you sorted all figures and tables into the sections of the text, please also sort the appendix figures and appendix tables into the respective appendix sections.
%% They will be correctly named automatically.

%% Option 2: If you put all figures after the reference list, please insert appendix tables and figures after the normal tables and figures.
%% To rename them correctly to A1, A2, etc., please add the following commands in front of them:

%\appendixfigures  %% needs to be added in front of appendix figures

%\appendixtables   %% needs to be added in front of appendix tables

%% Please add \clearpage between each table and/or figure. Further guidelines on figures and tables can be found below.

%\clearpage

\authorcontribution{HW initiated this study and designed the sensitivity experiments with input from the coauthors. HW conducted the EAMv0 simulation. SZ carried out the EAMv1 simulations and processed all the model output. HW and SZ led the analysis of the results and the other authors provided feedback. HW wrote the first draft of the manuscript and led the subsequent revisions. All coauthors contributed to the revisions.} %% this section is mandatory

\competinginterests{The authors declare no competing interests.} %% this section is mandatory even if you declare that no competing interests are present

%\disclaimer{TEXT} %% optional section

\begin{acknowledgements}
The authors thank Kai Zhang for helpful discussions during this study and 
for his comments on various versions of the paper. 
Dr. Andrew Barrett and an anonymous referee are thanked for their 
insightful reviews which helped to substantially improve the clarity of the paper.
This work was supported by the U.S. Department of Energy (DOE), 
Office of Science, Office of Biological and Environmental Research (BER)
%and Office of Advanced Scientific Computing Research (ASCR)
via the Scientific Discovery through Advanced Computing (SciDAC) program.
Computing resources were provided by the
National Energy Research Scientific Computing Center (NERSC), 
a DOE Office of Science User Facility operated 
under Contract No. DE-AC02-05CH11231.
Pacific Northwest National Laboratory is operated for DOE by Battelle Memorial Institute under contract DE-AC06-76RLO 1830.
\end{acknowledgements}

\end{document}